\documentclass{article}
\usepackage{url} 
\usepackage{tabularx}
\usepackage{ragged2e}
\usepackage{soul}
\usepackage{hyperref}
\usepackage{wrapfig} 

\usepackage[section]{algorithm}      
\usepackage[noend]{algpseudocode}    

\usepackage{amsmath}
\usepackage{graphicx}
\graphicspath{ {./figures/} }
\usepackage[colorinlistoftodos]{todonotes}
\usepackage{multirow}
\usepackage{booktabs}
\usepackage{caption}
\usepackage{tabularx}
\usepackage{xcolor}
\usepackage{gensymb}
\usepackage{float}
\usepackage[capitalize,nameinlink,noabbrev]{cleveref} 
\usepackage{gensymb} 
\usepackage{algorithm}    
\usepackage{placeins}
\usepackage{algpseudocode}    

\usepackage{tikz}
\usetikzlibrary{shapes,arrows,positioning}

\DeclareMathSizes{10}{9.5}{6}{6}
\makeatletter
\g@addto@macro \normalsize {%
 \setlength\abovedisplayskip{5pt plus 2pt minus 2pt}%
 \setlength\belowdisplayskip{5pt plus 2pt minus 2pt}%
}
\makeatother

\newcommand{\unk}{\mathbf{u}}
\newcommand{\fwd}{\mathcal{G}}
\newcommand{\pr}{p_0}
\newcommand{\llh}{L}
\newcommand{\pst}{p_L}
\newcommand{\prop}{\tilde{\unk}}
\usepackage{authblk}

\title{Methodological reconstruction of historical seismic events from anecdotal accounts of destructive tsunamis: a case study for the great 1852 Banda arc mega-thrust earthquake and tsunami}

 \author[1]{H. Ringer}
 \author[2]{J. P. Whitehead\footnote{correspondence to {whitehead@mathematics.byu.edu}}}
 \author[3]{J. Krometis}
 \author[4]{R. A. Harris}
 \author[5]{N. Glatt-Holtz}
 \author[6]{S. Giddens}
\author[4]{C. Ashcraft}
\author[2]{G. Carver}
\author[7]{A. Robertson}
\author[2]{M. Harward}
\author[2]{J. Fullwood}
\author[2]{K. Lightheart}
\author[2]{R. Hilton}
\author[2]{A. Avery}
\author[2]{C. Kesler}
\author[2]{M. Morrise}
\author[8]{M. H. Klein}

\affil[1]{Department of Mathematics, Virginia Tech University}
\affil[2]{Department of Mathematics, Brigham Young University}
\affil[3]{Advanced Research Computing, Virginia Tech University}
\affil[4]{Department of Geological Sciences, Brigham Young University}
\affil[5]{Department of Mathematics, Tulane University}
\affil[6]{Department of Mathematics, Notre Dame University}
\affil[7]{Department of Mathematics, Utah State University}
\affil[8]{Electrical and Computer Engineering, Duke University}

\begin{document}



\maketitle

\begin{abstract} 
  We demonstrate the efficacy of a Bayesian statistical inversion framework for reconstructing the likely characteristics of large pre-instrumentation earthquakes from historical records of tsunami observations.  Our framework is designed and implemented for the estimation of the location and magnitude of seismic events from anecdotal accounts of tsunamis including shoreline wave arrival times, heights, and inundation lengths over a variety of spatially separated observation locations.  The primary advantage of this approach is that all of the assumptions made in the inversion process are incorporated explicitly into the mathematical framework.  As an initial test case we use our framework to reconstruct the great 1852 earthquake and tsunami of eastern Indonesia.  Relying on the assumption that these observations were produced by a subducting thrust event, the posterior distribution indicates that the observables were the result of a massive mega-thrust event with magnitude near 8.8 Mw and a likely rupture zone in the north-eastern Banda arc.  The distribution of predicted epicentral locations overlaps with the largest major seismic gap in the region as indicated by instrumentally recorded seismic events. These results provide a geologic and seismic context for hazard risk assessment in coastal communities experiencing growing population and urbanization in Indonesia. In addition, the methodology demonstrated here highlights the potential for applying a Bayesian approach to enhance understanding of the seismic history of other subduction zones around the world.
\end{abstract}

\section{Introduction}
Indonesia is one of the most tectonically active and densely populated places on Earth. It is surrounded by subduction zones that accommodate the convergence of three of Earth's largest plates. Some of the largest earthquakes, tsunamis and volcanic eruptions known in world history happened in Indonesia \cite{Mc1999,harris2016waves}. Since these events, population and urbanization has increased exponentially in areas formerly destroyed by past geophysical hazards. Recurrence of some of these large events during the past two decades have claimed a quarter million lives \cite{NCEI_tsunami}.  

Most casualties from natural disasters in Indonesia are caused by tsunamis (the Indian Ocean earthquake and tsunami of 2004 is a prime example of this), which, over the past 400 years, occur on average every 3 years (e.g. \cite{hamzah2000tsunami}). Many potential tsunami source areas, such as the eastern Sunda  \cite{newcomb1987seismic} and Banda \cite{harris2011nature} subduction zones have no recorded mega-thrust earthquakes \cite{okal2003mechanism}.  However, some historical accounts of earthquakes and tsunamis in Indonesia provide enough detail about wave arrival times and wave heights from multiple locations to verify if mega-thrust events have happened in apparently quiescent regions, and assess the potential consequence of a similar event occurring in the future.  Indeed reliance on modern instrumental records of earthquake events to
determine seismic risk severely biases hazard assessments, as the
relevant temporal scales are hundreds or thousands of years on a given
fault zone. To improve risk estimates, it is imperative to draw from
historical records of damaging earthquakes, which reach beyond the fifty to seventy year horizon provided by modern instrumentation.

To this end, there has been substantial effort invested in the quantification of the characteristics of pre-instrumental earthquakes and tsunamis; see
e.g. \cite{newcomb1987seismic,sieh2008earthquake,meltzner2010coral,meltzner2012persistent,meltzner2015time, jankaew2008medieval,monecke20081,bondevik2008earth, bryant2007cosmogenic,grimes2006mapping,reid2016two,barkan2010tsunami,tanioka1996fault,nanayama2003unusually,LiuHarris2014, harris2016waves,fisherharris2016,GrNgCuCi2018,  martin2019reassessment}.  
As noted in these references, the historical and prehistorical data sources 
are sparse in details and laced with high levels of uncertainty. To improve 
the usage of these imprecise accounts, we develop a systematic framework to estimate earthquake parameters along with quantitative bounds on the uncertainty of these parameter estimates.  We do this using a Bayesian statistical inversion approach already leveraged in a variety of disciplines in the physical, social and engineering sciences, (see \cite{tarantola2005inverse,  kaipio2005statistical, dashti2017bayesian} as well as \cite{malinverno2002parsimonious,fukuda2008fully,sraj2014uncertainty,sraj2017quantifying}), to reconstruct large seismic events from historical accounts of the resulting tsunamis.  These efforts are related to a slew of recent and currently active work that seeks to determine the seismic source of modern tsunamis by inverting the available instrumental observations (see \cite{fujii2007tsunami,percival2011extraction,saito2011tsunami,giraldi2017bayesian,kubota2018tsunami,mulia2018adaptive} for example).

Our focus here is on an initial case study concerning the reconstruction of the 1852 Banda arc earthquake and tsunami in Indonesia detailed in the recently translated Wichmann catalog of earthquakes \cite{harris2016waves,wichmann1922earthquakes} and from contemporary newspaper accounts \cite{newspaper}.  To proceed with the Bayesian description of this inverse problem, we describe uncertainties in the noisy anecdotal  observations of the 1852 tsunami via probability distributions.  We next supplement this historical
data with a prior probability distribution for the seismic
parameters calibrated using modern instrumental seismic data. Finally, we develop a forward model mapping seismic
parameters to shoreline observations using  the Geoclaw software package
\cite{leveque2008high, leveque2011tsunami, gonzalez2011validation,
  berger2011geoclaw} to numerically integrate the nonlinear shallow water
equations, predicting the evolution of the tsunami initiated by seafloor deformation due to the earthquake itself.    These three elements are then combined with Bayes
theorem to produce a posterior 
distribution on the location, magnitude, and geometry of the most likely mega-thrust source for the 1852 tsunami.

Detailed information concerning the Bayesian posterior distribution, the output of our framework, is drawn from large scale computational simulations using Markov chain Monte Carlo sampling techniques \cite{liu2008monte}.  The solution of the inverse problem detailed here is reproducible from the described assumptions. Any of the assumptions can be modified by changing a few lines of code using a Python based software package available to the public upon request, and accessible via GitHub: \url{https://github.com/jpw37/tsunamibayes}.  This highlights the utility of this approach, in that the underlying assumptions are explicitly contained in the methodology, and as a result are inherently incorporated into the results, i.e. modification of the assumptions result in naturally occurring changes to the posterior distribution. Moreover, the Bayesian approach regularizes the inverse problem in the sense that small changes in the inputs should result in small changes to the results; see discussion in \cref{sec:discussion}. Allowing for a mega-thrust event as justified below, we find that the most likely cause of the observations for this 1852 account was an earthquake with magnitude near $8.8$ Mw and centroid to the south east of the island of Seram.

The rest of this article is organized as follows.  The next section includes a review of previous efforts related to this specific historical event, and a discussion of the source of the 1852 tsunami (mega-thrust earthquake or submarine slump).  Section 
\ref{sec:tectonic} describes the tectonic setting of the region in consideration.  Section \ref{sec:methodology} gives a very brief overview of the Bayesian methodology, a description of the different assumptions and parameterizations used for this particular event as well as an overview of the relevant historical observations and the forward tsunami model used here.  Section \ref{sec:Results} discusses the results of the inference and describes in some detail the posterior distribution that yields information on the possible earthquakes that may have resulted in the observed tsunami.  Finally Section \ref{sec:discussion} discusses the implications that can be derived from the posterior distribution, and a discussion of future work.

\section{Available observational data, and previous modeling for the 1852 Banda arc event}
For the Banda arc in particular, we note that there were two major tsunamis possibly connected to mega-thrust earthquakes in eastern Indonesia, witnessed in 1629 and 1852 \cite{wichmann1918earthquakes}. Numerical models of these events \cite{LiuHarris2014,fisherharris2016} show they were likely sourced from the shallow subduction interface of the Banda arc subduction zone \cite{harris2011nature}. This subduction zone  is largely ignored as a potential source for mega-thrust earthquakes because it involves the continental margin of northern Australia \cite{heuret2011physical,heuret2012relation,Cummins2020}. However, the high influx of accreted material from the continental margin in the Banda subduction zone overwhelms the capacity of the subduction channel in similar ways to the western Sunda arc of Sumatra where parts of the Bengal deep sea fan are subducting. Strong correlations exist between trench sediment thickness and megathrust earthquakes \cite{heuret2012relation}  The Sumatra region has experienced several mega-thrust earthquakes documented in instrumental, historical and geological records. One of the largest mega-thrust earthquakes in modern history happened along the northern Sunda arc in 2004. Geological records of tsunamigenic events from this region indicate average repeat times over the past 7400 years of around 450 years \cite{rubin2017highly}.  This statement is somewhat misleading though, as the temporal interval between events ranges from more than 2,000 years down to half a century. Such a variance in the temporal scales of seismic activity for a single region indicates that a lack of instrumental or even historical accounts of mega-thrust earthquakes on a given segment of a subduction zone should not be interpreted as unlikely, but rather as inevitable.  \cite{mccaffrey2007next} argues that no low angle convergent plate boundary with over 20 mm/a of convergence should be ruled out for producing mega-thrust earthquakes. Plate convergence across the Tanimbar and Seram Troughs varies from 30-70 mm/a \cite{bock2003crustal}.

The only large earthquake recorded instrumentally near the Tanimbar and Seram Troughs is a Mw $=$ 8.4 event in 1938 \cite{okal2003mechanism}. This earthquake was a widely felt thrust event, but the hypocenter was too deep (60 km) to cause a tsunami $> 1$ m \cite{anon1940}. Apparently the earthquake also did not cause a landslide induced tsunami even though there was intense shaking for several minutes.  On the other hand, historical accounts of the 1852 event are much more characteristic of a mega-thrust earthquake. Like the 1938 event, it was widely felt with an estimated MMI III minimum diameter $>$ 1100 km \cite{fisherharris2016}. Unlike the deep 1938 event however, the historical account from 1852 records a tsunami wave estimated at $>$ 8 m high in parts of the Banda arc \cite{fisherharris2016}.

Rather than considering a mega-thrust event as the source of the 1852 event, \cite{Cummins2020} hypothesize a submarine landslide as the primary source of the tsunami, with the latitude-longitude location of the landslide near where \cite{fisherharris2016} found from limited tsunami modeling was the best-fit centroid of a mega-thrust earthquake. There are two primary reasons why a mega-thrust earthquake can not be discounted versus a submarine landslide caused by a smaller earthquake as hypothesized by \cite{Cummins2020}.  Let us now successively describe both of these reasons.
 
 First, we note that it is well established that landslide induced tsunamis, while locally more forceful and with much higher run-up wave heights, typically have a significantly shorter wave-length from the initial seafloor disturbance and hence dissipate much more rapidly than a seismically induced tsunami.  With this in mind, \cite{OkalSynolakis2004} use experimental data to develop a quantitative heuristic to assist in determining the relative likelihood of a given tsunami being induced by a landslide or directly by the earthquake.  They define $I_2$ as the ratio between the maximum run-up wave height, and the horizontal extent of the wave run-up, and show that $I_2 < 10^{-4}$ for tsunamis induced by a pure seismic event, and $I_2 > 10^{-4}$ for tsunamis induced by a landslide. Although \cite{OkalSynolakis2004} assumes a single, straight shoreline perpendicular to the wave, the heuristic is shown to be remarkably accurate even for more realistic geometries.  For the current setting, even if we presume a maximal run-up wave height of $8m$ based on the historical account at Banda Neira, then $I_2 = \frac{8m}{340 km} \approx 2.35 \times 10^{-5}$ which is right in line with the values calculated by \cite{OkalSynolakis2004} for seismically induced tsunamis, but significantly less than that for landslide induced ones. In summary the lateral reach of the tsunami itself, evident from the historical account for the 1852 event, is far too broad to warrant a landslide induced source.  
 
 Second, \cite{Cummins2020} rightly emphasize that the eastward convergence of a thrust earthquake along the west-dipping subduction interface would produce a negative wave in the Banda Sea prior to the arrival of the first positive wave, something that is not recorded in any of the historical records, but shown in numerical models \cite{fisherharris2016} including those utilized below.  While this argument is certainly accurate, we note that the absence of such information in the historical account does not necessarily preclude the occurrence of such a negative wave. The morning of the event in question was a spring tide which caused an extremely low tide level so that a small negative wave would be far less noticeable. In addition, in the 1850's the Banda Sea region observed Moluccan time, which in November would indicate a sunrise at approximately 7:30AM, only minutes prior to the earthquake.  It is very likely that the Dutch officers (the primary source of the historical record) would not take notice of a weak negative wave so near sunrise at a spring low-tide, particularly following a devastating earthquake.

In addition to these two primary reasons, we note that the 1938 Mw = 8.4 earthquake \cite{okal2003mechanism} was very nearly in the same location as the proposed landslide source in \cite{Cummins2020}, yet there is no evidence of a submarine slump occurring as a result.  This makes it less likely that the earthquake source proposed in \cite{Cummins2020} may have induced a slump sufficient to yield the recorded tsunami.

These arguments do not eliminate the possibility that the tsunami in question was caused by submarine slumping along the edge of the Weber Deep as proposed by \cite{Cummins2020}, but it does indicate that the potential for a mega-thrust event as the primary source is more likely and therefore can not be disregarded.  This paper addresses the mega-thrust hypothesis directly and systematically  shows that a  mega-thrust source along the Tanimbar and Seram troughs can produce a tsunami that matches the historical account.  In addition, the systematic approach taken here clearly describes the most likely location, strength and geometric layout of the 1852 earthquake if it was indeed a mega-thrust event.



\section{Tectonic Setting}\label{sec:tectonic}

\begin{figure}[h]
    \centering
    \includegraphics[width=.8\textwidth]{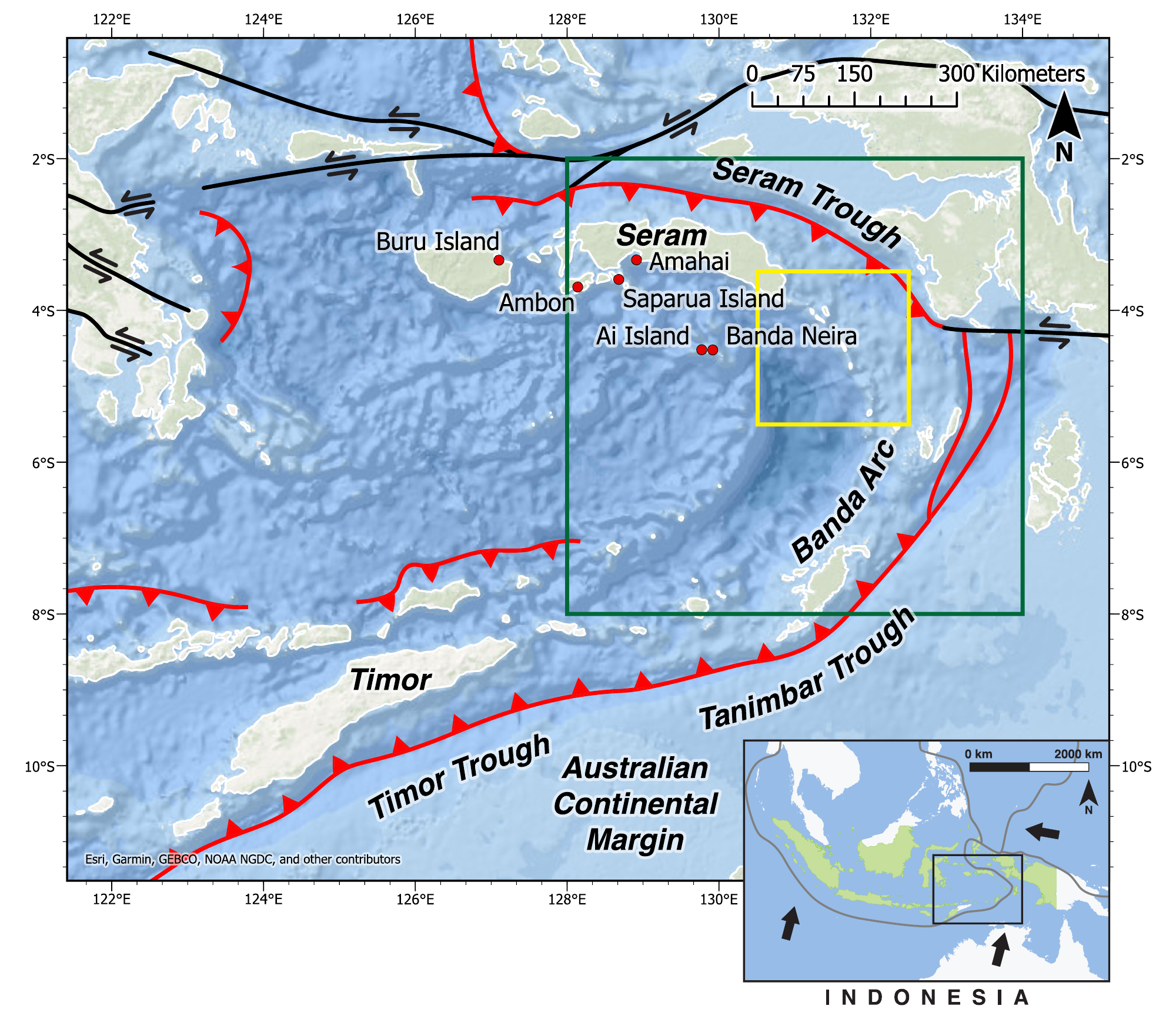}
    \caption[1852 Banda arc event observation locations]{The seismic and geologic setting for the 1852 event.  The convergent plate boundaries are in red, the transvergent boundaries are in black.  The arrows in the inset map indicate motion of the Pacific and Australian plates relative to the Eurasian plate.  The Australian plate is converging at a rate of 70mm/yr and the Pacific plate is converging at 110mm/yr. The nine observation locations from the Wichmann catalog for the 1852 Banda arc earthquake and tsunami are also labeled.  The green rectangle indicates the region that is depicted in \cref{fig:latlonpost} and \cref{fig:seismic-gap}.  The yellow rectangle is the region depicted in \cref{fig:samp_cond}.}
    \label{fig:obslocations}
\end{figure}

The Banda arc subduction zone is the eastward extension of the Sunda arc where both arcs rise above where the Australian Plate subducts beneath the Eurasian Plate. The composition of the lower and upper plate reverses at the boundary between the Sunda and Banda arcs: the lower plate changes eastward from oceanic to continental and the reverse happens to  the upper plate. Both arcs are partial subduction zones, and partial collision zones.  The western Sunda arc is a collision between deep sea fan deposits riding on an oceanic lower plate and a continental arc upper plate, while the Banda arc is a collision between passive margin deposits of NW Australia and an oceanic arc \cite{hamilton1979tectonics}. At both subduction zones the Australian plate moves NNE relative to the Eurasian plate.  However, the rates of motion across the subduction zone in the Banda arc (70 mm/a, \cite{nugroho2009plate}) are nearly twice those of the western Sunda arc \cite{bock2003crustal}. 

Like the western Sunda arc, the Banda arc consists of two chains of mountains, an inner volcanic arc and an offshore chain of rising islands associated with offscraping and accretion of thick layers of mostly sedimentary rock riding on the subducting plate \cite{hamilton1979tectonics}. In the Banda arc the accreted layers are part of the distal Australian passive continental margin \cite{carter1976stratigraphical,harris1991temporal}, which according to multiple sources of data, has subducted to at least a depth of 300 km \cite{harris2011nature,tate2015australia}. The Banda volcanic arc is still active though it is largely contaminated by subducted Australian continental crust \cite{whitford1977geochemistry,hilton1989helium}.  What was a deep trench before the buoyant continental lithosphere arrived at the plate boundary is now a `trough' known from SW to NE as the Timor, Tanimbar and Seram troughs, see \cref{fig:obslocations}. Both subduction zone interfaces are active, and rupture through sedimentary layers that drape over the subduction interface. Earthquakes along this interface yield fault-plane solutions with low-angle thrusts and mostly dip-slip slip vectors. Shortening throughout the islands rising above these active thrusts also verge perpendicular to the plate boundary \cite{harris2011nature}. 

Another similarity between the Sumatra section of the Sunda arc and the Banda arc is that each changes considerably in strike, which causes highly oblique plate convergence in some areas. Notwithstanding this obliquity, several mega-thrust earthquakes are recorded in the western Sunda arc. Slip rakes of these events are mostly parallel to one another, but largely perpendicular to the trench at the epicenter, which is nearly 90 degrees from the Australian plate convergence direction.  

We challenge the idea based on geological, geophysical and historical data that the eastern Sunda arc and Banda arc can be dismissed as potential sources of mega-thrust earthquakes and tsunamis. In addition, the primary results of this study, illustrated in \cref{fig:latlonpost} and \cref{fig:priorvspost}, imply that the 1852 event was most probably located along a narrow region in the eastern portion of the Banda arc and was a massive mega-thrust earthquake on the same scale as the December 2004 Sumatra event.

\section{Methodology}\label{sec:methodology}

This section lays out the complete details of the methodology we use to develop our Bayesian statistical model of the 1852 event. We start with an overview of the components of the model.

\subsection{Overview of the Bayesian Statistical Model}

The Bayesian approach to statistical inversion,
cf. \cite{gelman2014bayesian, kaipio2005statistical,
  dashti2017bayesian, tarantola2005inverse}, provides a methodology
for converting uncertain outputs of a physical model into
probabilistic estimates of model parameters.  This framework is
perfectly suited for the anecdotal, uncertain nature of the historical
accounts utilized here.  Using this Bayesian methodology provides a
lot of flexibility and will be adjusted to treat a variety of other
tsunami producing seismic events from historical and pre-historical
data sets in future studies.

The primary inputs required for Bayesian statistical inversion, particularly applied to the determination of historical earthquakes are: 
\begin{itemize}
\item[(1)] A \emph{prior probability distribution} $\pr$ describing the best
  guess of a set of earthquake parameters $\unk$ without considering the observations. For the 1852 event, we formulate the prior
  distribution via independent distributions on the depth (and hence location) and the magnitude (and hence length and width of the rupture zone).  Further details as well as a description of the geometric layout of the ruptures are provided
  in \cref{sec:prior:consts}.
\item[(2)] A \emph{likelihood probability distribution} $\llh$ describing measured data and observational uncertainties,
  which for the 1852 event are observations of tsunami arrival time,
  height, and coastal inundation taken from the Wichmann catalog.  The
  associated uncertainties are estimated from a direct textual
  analysis combined with other information about the shoreline
  locations where the event was recorded. \cref{sec:His:data}
  describes the selection of this dataset.
\item[(3)] A \emph{forward model} $\fwd$ describing the relationship between
  model parameters and observations. For the 1852 event, we use the
  Geoclaw software package to propagate off-shore tsunami waves, supplemented with a heuristic model that maps on shore wave heights and shoreline geometry to inundation length.  See \cref{sec:forward}
  for additional details.
\end{itemize}

With these inputs specified, Bayes' Theorem gives the \emph{posterior}
probability distribution $\pst$, i.e.  the conditional distribution of
$\unk$ given $\llh(\fwd(\unk))$:
\begin{align}
  \label{eq:post}
  \pst(\unk) 
  = \frac{1}{Z} \llh(\fwd(\unk)) \pr(\unk)
\end{align}
where $Z := \int \llh(\fwd(\unk)) \pr(\unk) d\unk$ is the normalizing
constant. The posterior describes in probabilistic terms the
seismic parameters $\unk$ that best match both our understanding of
reasonable parameter values based on the prior distribution \emph{and}
observations extracted from historical records associated with the
1852 Banda arc event. In order to extract quantitative information from our model, including a variety of marginal distributions correlating variables of interest, we make use of Markov
chain Monte Carlo (MCMC) statistical sampling techniques
\cite{kaipio2005statistical,liu2008monte}. See \cref{sec:mcmc} for details.

\subsection{Calibrating the parameter space 
  and the prior distribution}
\label{sec:prior:consts}

To make efficient use of Bayesian methods, it is necessary to consider the dimensionality of the parameter space. As the number of parameters to be estimated increases, so does the difficulty of the sampling problem. This `curse of dimensionality' appears in this setting because Bayesian inference boils down to the computation of high dimensional integrals. It is known that random walk MCMC methods converge arbitrarily slowly as the dimension of the parameter space increases \cite{roberts1997weak,roberts1998optimal,beskos2009optimal}.  Thus, we need to ensure that the dimension of the parameter space we use to describe the earthquakes remains relatively `small'.

\begin{figure}
    \centering
    \includegraphics[width=.9\textwidth]{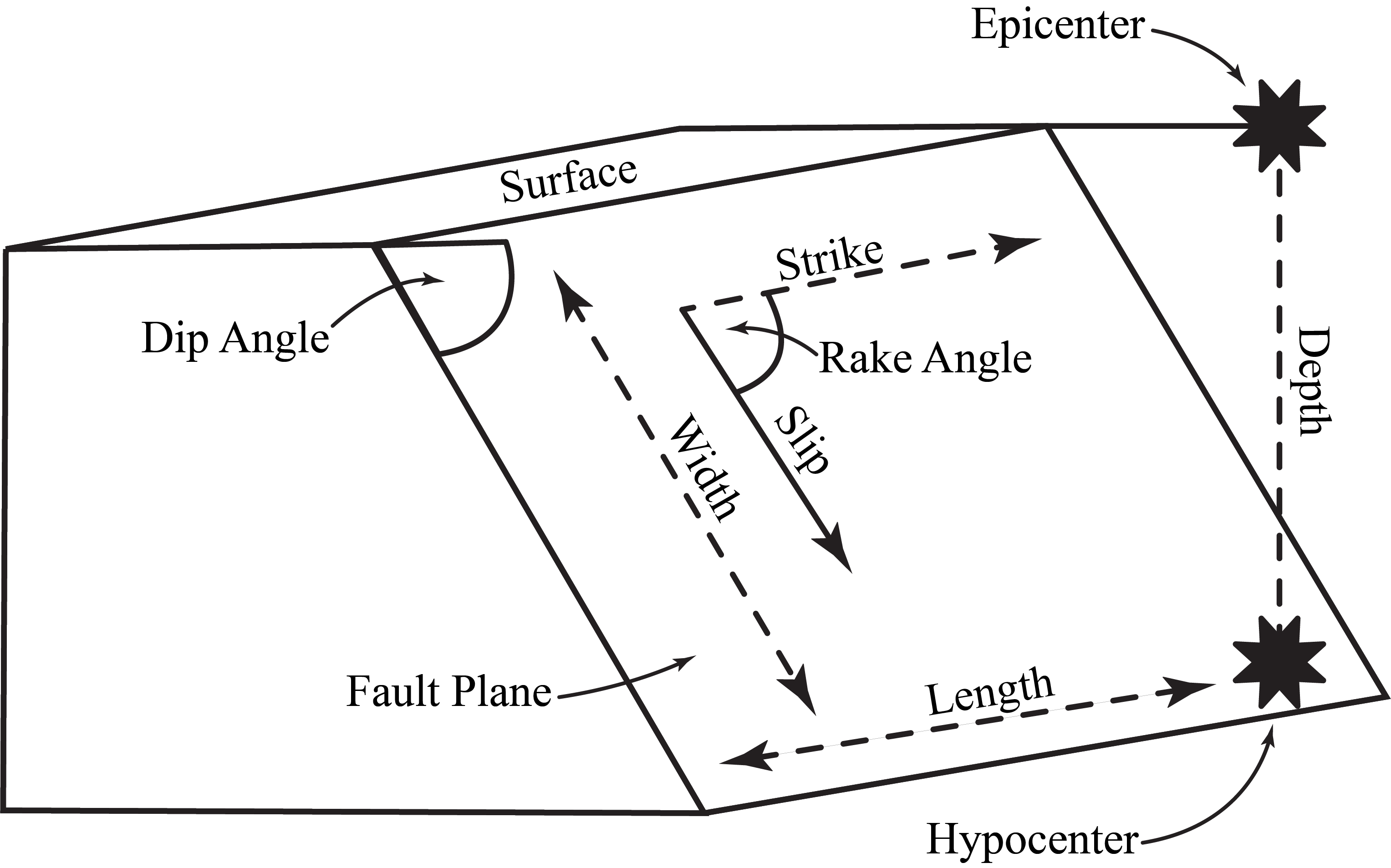}
    \caption[Okada demonstration]{A sketch of the fault plane geometry illustrating the Okada parameters for an instantaneous rupture.  The strike angle is the direction parallel to the length of the fault, which is measured in latitude-longitude relative to North.  Dip indicates the angle from horizontal the fault is inclined, and rake is the angle that indicates in what direction the slip vector (the fault rupture) moves relative to strike.  The other Okada parameters are the hypocentral or epicentral location as specified by latitude-longitude coordinates, the depth of the hypocenter, the amount of slip, and the length and width of the rectangular fault.}
    \label{fig:Okada}
\end{figure}

A zeroth order approach to parameterizing an earthquake, is to consider the 9-dimensional parameter space for the Okada model \cite{okada1985surface,okada1992internal} for an instantaneous rupture, as illustrated in \cref{fig:Okada}.  In the current study, we do not consider a time-dependent rupture, although we will pursue such considerations in future work.  Even with the assumption of an instantaneous rupture, it is unreasonable to model the source earthquake as a single rectangular rupture along the Banda arc where the strike changes quite rapidly with latitude and longitude. Naively, an alternative would be to create an $N$-subfault rupture zone made up of $N$ rectangular subfaults that follow the geometry of the subduction zone, allowing for realistic changes in the strike.  Such a model would require a $9N$-dimensional parameter space, which produces an intractable sampling problem for any useful value of $N$ (to capture the curvature of the Banda arc, we need a minimum of $N \geq 3$).

To reduce the dimensionality of the parameter space, we make a distinction between the \textit{(forward) model parameters} and the \textit{sample parameters}. The model parameters are the direct inputs to the forward model: in this case, the $9N$ Okada parameters for an $N$-subfault rupture. On the other hand the sample parameters are a reduced representation of the model parameters.  As such this lower dimensional subset of sample parameters is where we define our posterior distribution and therefore determines the dimension of the space where the MCMC is carried out. 

For reducing the number of sample parameters, a good starting point is to consider model parameters that may be assumed to take constant values. Among the nine Okada parameters, the rake angle can be reasonably fixed to 90\degree. This corresponds to pure thrust motion, which acts perpendicular to the strike of the fault. While strike-slip motion is certainly present in real mega-thrust earthquakes, thrust motion is the primary driver of seafloor deformation, and thus tsunami formation. Within the Okada model, rake angles other than $\theta=90\degree$ are roughly equivalent to a reduction in slip by a factor of $\sin(\theta)$. This coupling between the slip and rake angle make it nearly impossible to distinguish the effect of one over the other in the inference, so we leave the rake fixed throughout, and infer the slip instead.

Another avenue to reduce the dimensionality is to seek model parameters that can be determined from other model parameters in the context of prior information. For the 1852 Banda arc earthquake, a detailed model of the subduction zone geometry is available from the USGS \emph{Slab2} dataset \cite{HayesSlab2}. 
Depth, dip angle, and strike angle can be determined from latitude and longitude, although each of these parameters has inherent uncertainties that are associated with each variable.


Fixing the rake angle, and determining depth, strike and dip from latitude-longitude, we are left with five of the Okada parameters: latitude, longitude, length, width, and slip. These could be chosen as the sample parameter space. However, a problem arises in choosing the triple of (length, width, slip) as sample parameters, due to their relationship with earthquake magnitude. The \emph{scalar seismic moment} $M_0$ of an earthquake of length $L$, width $W$, and average slip $S$ is defined as
\begin{equation}
M_0 = \mu LWS,
\label{eq:seismoment}
\end{equation}
where $\mu$ is the shear modulus of the rock (or Earth's crust), with dimensions of force per unit area. 
Using this definition, the \emph{moment magnitude} Mw \cite{kanamori1979,USGSmaggroup} is defined as
\begin{equation}
\mbox{Mw} = \frac{2}{3}(\log_{10} M_0-9.05).
\label{eq:momag}
\end{equation}
It is clear that the empirical frequency of earthquakes of a given moment magnitude follows an exponential distribution \cite{Kagan2002}, i.e. smaller earthquakes are exponentially more likely to occur than large magnitude earthquakes. In order to ensure that magnitude follows an exponential prior distribution, we remove slip from the sample parameters and replace it with moment magnitude. Given values of (moment) magnitude, length, and width, slip can be back-calculated via Equations \ref{eq:seismoment} and \ref{eq:momag}.

\begin{figure}
    \centering
    \includegraphics[width=.9\textwidth]{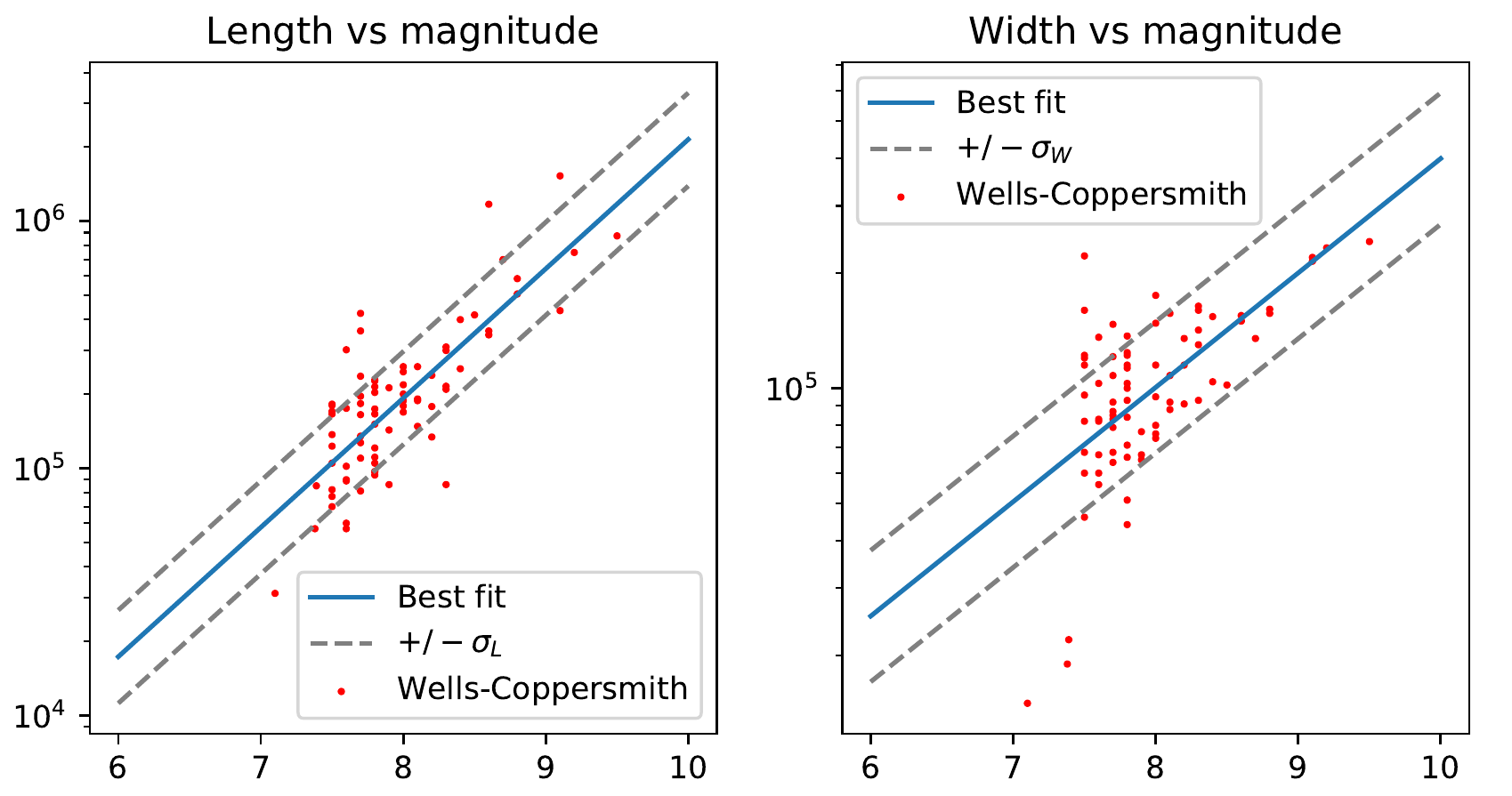}
    \caption[Magnitude-normalized length and width best fits]{Wells-Coppersmith and modern data plotted alongside linear best fits for $\log L$ and $\log W$ against Mw.}
    \label{fig:lengthwidth}
\end{figure}

Equations \ref{eq:seismoment} and \ref{eq:momag} also highlight a challenge when using a random walk proposal kernel (something we are restricted to because the forward model is far too complicated for any type of gradient based method and the non-Gaussianity of the prior eliminates all other options) with these parameters. Since magnitude grows with the logarithm of length and width, any fixed choice of variance for length and width in the Gaussian proposal kernel will be inappropriate for all but a limited range of magnitudes. Therefore, we introduce magnitude-normalized substitutes for length and width as sample parameters. Using the Wells-Coppersmith dataset \cite{wells1994new} (adjusted with additional data collected for more recent and mega-thrust level earthquakes), we computed linear least squares fits for $\log L$ and $\log W$ against magnitude. These fits are displayed in Figure \ref{fig:lengthwidth}. Our magnitude-normalized substitutes for length and width are $\Delta \log L$ and $\Delta \log W$, the ``residuals'' compared to the linear best fit. In other words: given values for Mw, $\Delta \log L$, and $\Delta \log W$, length and width are computed as:
\begin{align*}
    \log L &= a\mbox{Mw}+b+\Delta \log L \\
    \log W &= c\mbox{Mw}+d+\Delta \log W
\end{align*}
where $a,b,c,d$ are the coefficients of the linear best fits as shown in Figure \ref{fig:lengthwidth}.  Once the magnitude Mw, and length and width are found, then the slip can be computed from \eqref{eq:seismoment} and \eqref{eq:momag}.

To the five sample parameters (latitude, longitude, magnitude, $\Delta \log L$, $\Delta \log W$), we add a sixth parameter: \emph{depth offset}. The Slab2 data includes estimates of uncertainty in the subduction interface depth
. Depth offset accounts for this uncertainty by allowing for earthquakes that are situated somewhat deeper or shallower than is specified in the Slab2 depth map.  By the above approach, we reduce dozens or hundreds of Okada parameters to a set of six -- latitude, longitude, magnitude, $\Delta \log L$, $\Delta \log W$, and depth offset -- that is both low enough in dimension to be computationally tractable and sufficiently independent to well-represent the possible earthquakes, as we describe next.

\subsubsection{Computing Subfault Model Parameters}

As mentioned above, it is necessary to model the earthquake as a collection of rectangular subfaults that conform to the subduction interface geometry. Here we describe our approach for ``decompressing'' the six sample parameters introduced above into the Okada parameters for N rectangular subfaults that follow the interface geometry provided by Slab2.

The basic approach is to ``break'' a single rectangular rupture zone into an $m \times n$ grid of identical subrectangles, which are then oriented to conform to the interface geometry. Each of these subrectangles has length $L/m$ and width $W/n$, where $L$ and $W$ are the length and width of the full rupture zone.  The difficulty then lies in choosing the number of subfaults, and identifying the dip angle and strike orientation for each one so as to best match the fault geometry.

\begin{figure}[h]
    \centering
    \includegraphics[width=.75\textwidth]{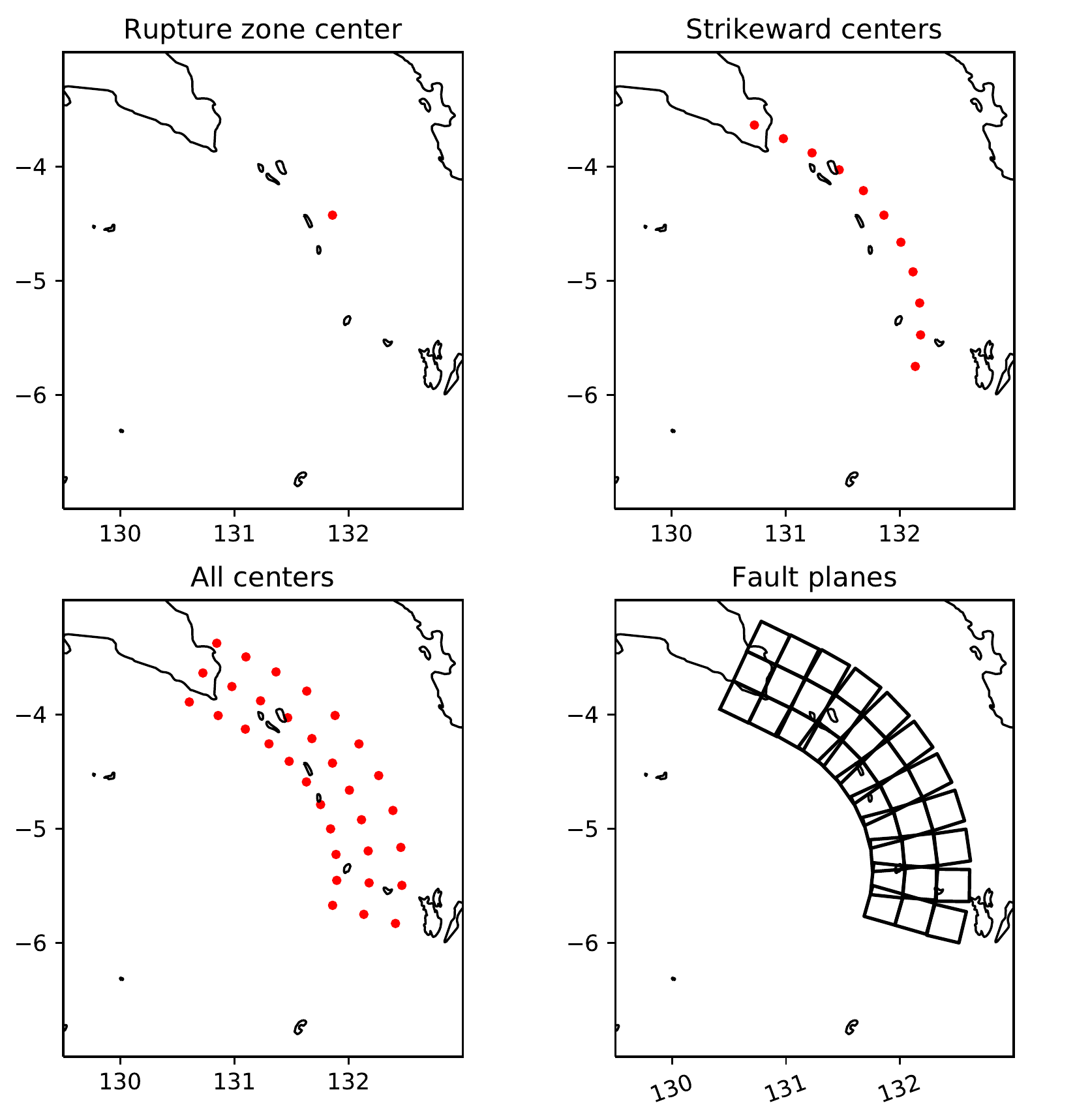}
    \caption[Rectangular subfault grid generation]{Placing subrectangles contoured to interface geometry. First, a point is placed at the center of the rupture zone. Points are then placed forwards and backwards following the strike angle (essentially following level curves of depth). Additional points are placed up-dip and down-dip. Using Slab2 depth, dip, and strike data, Okada parameters for rectangles centered at each point are computed.}
    \label{fig:subfault_split}
\end{figure}

For ease of implementation, we use odd values of $m$ and $n$. We found that $m=11$ and $n=3$ made an appropriate fit for the 1852 Banda arc earthquake, adequately capturing the geometry without over-fitting the parameter space.  These values of $m$ and $n$ are used for every sample, regardless of the magnitude.  To determine the orientation and location of each subfault, we place a single point at the latitude and longitude of the centroid of the full rupture zone. Using the Slab2 map of strike angle, we move in opposite directions, staying parallel to strike. Every $L/m$ kilometers, we place another point. This continues until $m$ points are placed. For each point, we then move in opposite directions, perpendicular to the strike angle, placing points every $W/n$ kilometers, until all $mn$ points have been placed. These points are the latitude/longitude coordinates for the centers of the subrectangles. This procedure is displayed in Figure \ref{fig:subfault_split}.

Having specified the latitude, longtitude, length, and width for each subrectangle, the remaining Okada parameters are determined as follows. Each subrectangle is given the same slip value as determined by \eqref{eq:seismoment} and \eqref{eq:momag}. The strike and dip angles are determined by the Slab2 strike and dip maps using the centroid of each subrectangle as the reference point. The depth is determined by the Slab2 depth map, plus the value of the depth offset sample parameter (universal to all subfaults). As discussed above, all subfaults are assigned a rake angle of 90\degree.

\FloatBarrier
\subsubsection{Specifying the full prior distributions}

Selection of appropriate prior distributions is a key step in good Bayesian inference. An over-specified prior can overwhelm the data, and an under-specified prior may allow for parameter values that are non-physical. Having discussed the map from sample parameters to model parameters, we now discuss our choice of prior distributions for latitude, longitude, magnitude, $\Delta \log L$, $\Delta \log W$, and depth offset for the 1852 Banda arc event.

Prior constraints on earthquake latitude and longitude are derived from the subduction interface geometry. Large earthquakes can only be supported in a certain range of depth: too deep, and the crust is too plastic to store the elastic strain energy necessary for a large earthquake \cite{sallares2019}, to shallow, and the rupture interface would extend above the surface. In addition, an earthquake with the potential for generating a tsunami is even more constrained in depth.  We take the approach that, a priori, depth is the primary constraint on earthquake location. Since the Slab2 dataset gives a depth map for the subduction zone along the entire Banda arc, any probability distribution on depth produces an implied distribution on latitude and longitude, at least for mega-thrust events like those considered here. Based on the augmented Wells-Coppersmith dataset, we chose a truncated normal distribution for depth. This distribution is supported on $[2.5,50]$ kilometers, with a mean of 30km and a standard deviation of 5km. Evalutating the pdf of this distribution at each latitude/longtitude coordinate, via the Slab2 depth map, gives a non-negative continuous function. Although this function does not integrate to unity, the normalizing constant cancels out in the evaluation of the Metropolis-Hastings acceptance parameter $\alpha$, i.e. this function provides an adequate weight for the latitude and longitude sample parameters. The unnormalized logpdf of the latitude/longitude prior is displayed in the left panel of \cref{fig:latlonpost}.  As this Figure illustrates the likely locations of the centroid of the earthquake only, there is potential for an event that extends above the surface, if the centroid is located near $2.5km$ in depth.  These events are automatically rejected in the Metropolis-Hastings step by assigning a likelihood of $0$.


As discussed above, earthquake magnitude is observed to approximately follow an exponential distribution. Clearly however, the exponential scaling cannot continue indefinitely in the large magnitude regime, and a number of approaches have been used to address this (see \cite{Kagan2002}). We take the simple approach of right-truncating the exponential distribution at magnitude 9.5, i.e. we do not allow for an event of magnitude greater than $9.5$ Mw.  A consensus estimate for the parameter of the exponential distribution is $\lambda = .5$ \cite{Kagan2002}, which we use here.

Since $\Delta \log L$ and $\Delta \log W$ are magnitude-normalized length and width, defined as residuals against a linear best-fit, we chose Gaussian prior distributions with mean zero for each of these sample parameters. The standard deviations for these distributions are determined from the sample variances for the residuals in the augmented Wells-Coppersmith dataset against the linear fit (see Figure \ref{fig:lengthwidth}). These values are $\sigma_{\Delta \log L} = 0.188$ and $\sigma_{\Delta \log W} = 0.172$.

The prior for depth offset was chosen based on the Slab2 depth uncertainty data. The average reported uncertainty is roughly 5km, so a mean-zero normal distribution with standard deviation of 5 was selected.


\subsection{The Historical Dataset, the relevant uncertainties, and assigned likelihood distributions.}
\label{sec:His:data}
\subsubsection{Overview of historical account and potential observations}
Observations are selected from the historical accounts in the Wichmann catalog \cite{wichmann1918earthquakes, wichmann1922earthquakes} based on two key criteria.  First, the account has to provide an identifiable location (latitude-longitude) that can be incorporated into the modeling.  In other words, the details provided in the historical account must be sufficiently accurate to yield a precise location via modern-day maps and information.  Second, the account has to be sufficiently detailed that some level of confidence can be placed on the observable in question.  Note that drawing from a catalog of this kind introduces unavoidable ambiguities that do not apply to modern instrumental data.  For example, we specify the wave height based on passages of the form ``[t]he water rose to the roofs of the storehouses and homes,'' as described in more detail below.

Thirteen different observations for the 1852 Banda arc tsunami meet these criteria spread across nine locations, which are shown in Figure \ref{fig:obslocations}. These include three types of observations:
\begin{enumerate}
\item \emph{Arrival time.}  The arrival of the first significant wave after the shaking stopped.  We assume that the arrival time refers to the first wave, not the maximal one.
\item \emph{Maximum wave height.}  This is the most frequent
  observable, and is identified at every location.  We emphasize here that the observations are often taken near the shore, but not exactly on the shoreline.  To account for this, we widen the relevant likelihood distributions, although neglecting the shoreline to on-shore effects will only under-estimate the power of the wave and hence strength of the seismic event.
\item \emph{Inundation length.}  This refers to the distance inland that the wave traveled onshore, and is actually interpreted for our purposes as a deterministic function of the wave height.  This essentially places a double amount of weight on those locations that have observations of both wave height and inundation.
\end{enumerate}

Based on the text of each account, a probability distribution is developed describing the probabilistic likelihood that each observation took a given value. These distributions, which are assumed to be independent, are shown in \cref{fig:likelihoods}.  Rather than explain the reasoning behind all thirteen of these likelihood distributions for each of the nine locations, we only provide a detailed discussion of the likelihood for a single location: Banda Neira.


\subsubsection{Banda Neira: a sample likelihood distribution}

Observations at this location are primarily taken from page 242 in the Wichmann catalog which states in part: ``Barely had the ground been calm for a quarter of an hour when the flood wave crashed in...The water rose to the roofs of the storehouses and homes...[the wave] reached the base of the hill on which Fort Belgica is built on Banda Neira''.  We expect the wave height observation to be near the boat dock on Banda Neira which is just east of Fort Nassau.  For the available bathymetry data, we seek a location near that point that will maintain a sizable wave for a reasonably initiated tsunami.  With this in mind, we select $-4.524^\circ$ latitude and $129.8965^\circ$ longitude as the observation location.
    
Using 15 minutes as the anticipated arrival time of the wave at Banda Neira is too simplistic for these circumstances.  In particular it is noted in other locations that the shaking lasted for at least 5 minutes, but the modified Okada model used in Geoclaw here assumes an instantaneous rupture. Hence we build into the likelihood, a skew toward longer times with a mean of 15 minutes. This is done with a skew-normal distribution with a mean of 15 minutes, standard deviation of 5 minutes, and skew parameter 2.
    
Assuming standard construction for the time period for the homes (and storehouses) we can assume the water rose at least 4 meters above standard flood levels as most buildings of the time were built on stilts and had steep vaulted roofs.  Based on the regular storm activity in the region we can expect that with high tide, and normal seasonal storm surge, the standard flood level is also approximately 2 meters in this region.  This leads us to select a normally distributed likelihood for wave height with a mean of $6.5m$ and standard deviation of $1.5m$, allowing for reasonable likelihood values for wave heights in the range from $3m$ to $9m$.
    
To quantify the wave reaching the base of the hill, we measured the distance from 20 randomly selected points along the beach to the edge of said hill in arcGIS.  The mean of these measurements was 185 meters, with a standard deviation of roughly 65 meters. Thus we choose a normal distribution with those parameters.  Without more detailed information about the coastline, and a direct idea of the direction of the traveling wave, we can not be more precise with regard to the inundation, but this is sufficient for the model we use (as described below).

\subsubsection{Overview of all likelihoods}
\begin{figure}[H]
    \centering
    \includegraphics[width=\textwidth]{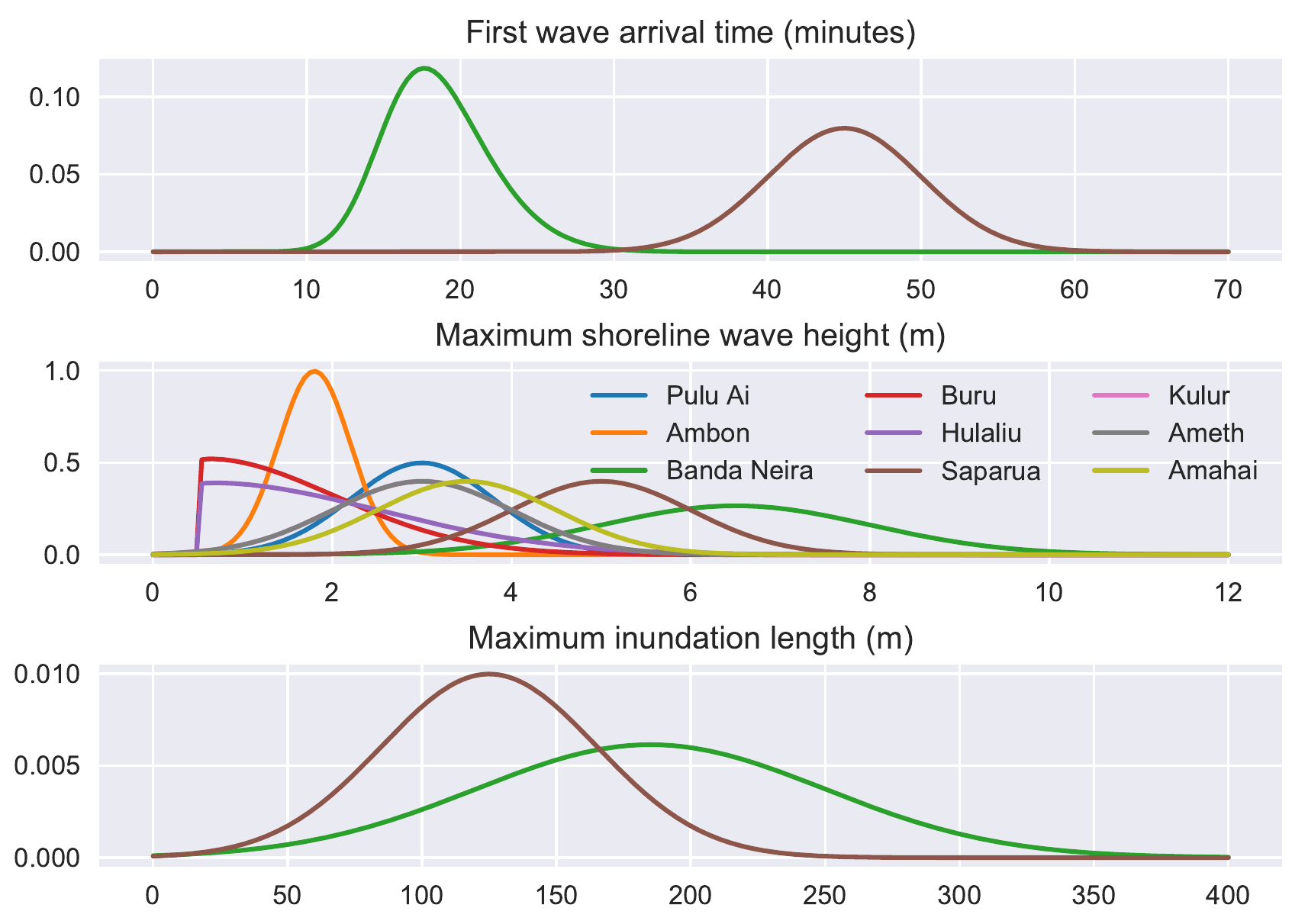}
    \caption[1852 Banda arc event likelihood densities]{1852 Banda arc tsunami likelihood densities for the 13 observations at 9 locations. Each likelihood density represents an interpretation of the Wichmann catalog description.  The same color scheme is used for all 3 types of observations, but only Banda Neira and Saparua included wave arrival time and inundation length.}
    \label{fig:likelihoods}
\end{figure}

The likelihood distributions for the other 8 locations are constructed in a very similar manner to that described above for Banda Neira 
and shown graphically in \cref{fig:likelihoods}.  The current investigation has assigned each likelihood to a single latitude-longitude location based on the historical record.  Such a specific assignment is reasonable only if the likelihood distributions are sufficiently wide to account for bathymetric and model dependent resolution differences along the coastline which is a reasonable assumption although certainly not one that is guaranteed.  In future studies we will address this issue by weighting the wave heights and arrival times from a collection of nearby latitude-longitude locations.

The total likelihood of a given event is computed as the product of these individual observational likelihoods (we rely heavily on the assumption that each observable is independent of the others).  The assumption of independence of the different observations is certainly questionable, but there is also no reason to suppose that a more complicated construction of the total likelihood is preferable, i.e. we have chosen to take the most simplified approach without making additional unjustifiable assumptions about the structure of the likelihood.

\subsection{The Forward Model}
\label{sec:forward}

Calculation of the forward model that maps seismic events to
quantitative observables is the most complicated and computationally
expensive part of the inversion process.  We compute the tsunami
observations resulting from seismic events by numerically integrating
the shallow water equations in a restricted region surrounding the Banda Sea.

\subsubsection{Geoclaw Integration and `high' resolution bathymetry}
\label{sec:forward:geo}
The propagation of the tsunami waves is computed via the nonlinear shallow water
equations supplemented with the appropriate initial and boundary
conditions dictated by the specified Okada parameters and bathymetry
of the region.  We simulate the tsunami generated by each Monte Carlo
sample using the Geoclaw software package, \cite{leveque2008high,
  leveque2011tsunami, gonzalez2011validation, berger2011geoclaw} which
employs an adaptively-generated mesh for a finite volume based scheme.
For bathymetry (sea-floor topography) we use the 1-arcminute etopo
datasets available from the open access NOAA database (
  \url{https://www.ngdc.noaa.gov/mgg/global/global.html}) referred to
hereafter as NOAA bathymetry, and for the coastline near each
observational point we utilize higher resolution Digital Elevation
Models (DEM) from the Consortiom for Spatial Information (CGIAR-CSI, \url{http://srtm.csi.cgiar.org/srtmdata/})
referred to below as DEM coastlines.  These higher resolution
topographical files yield a 3-arcsecond resolution on land, but give
no additional information on the sub-surface bathymetry. 

\begin{figure}[!htbp]
\centering
  \includegraphics[width=\textwidth]{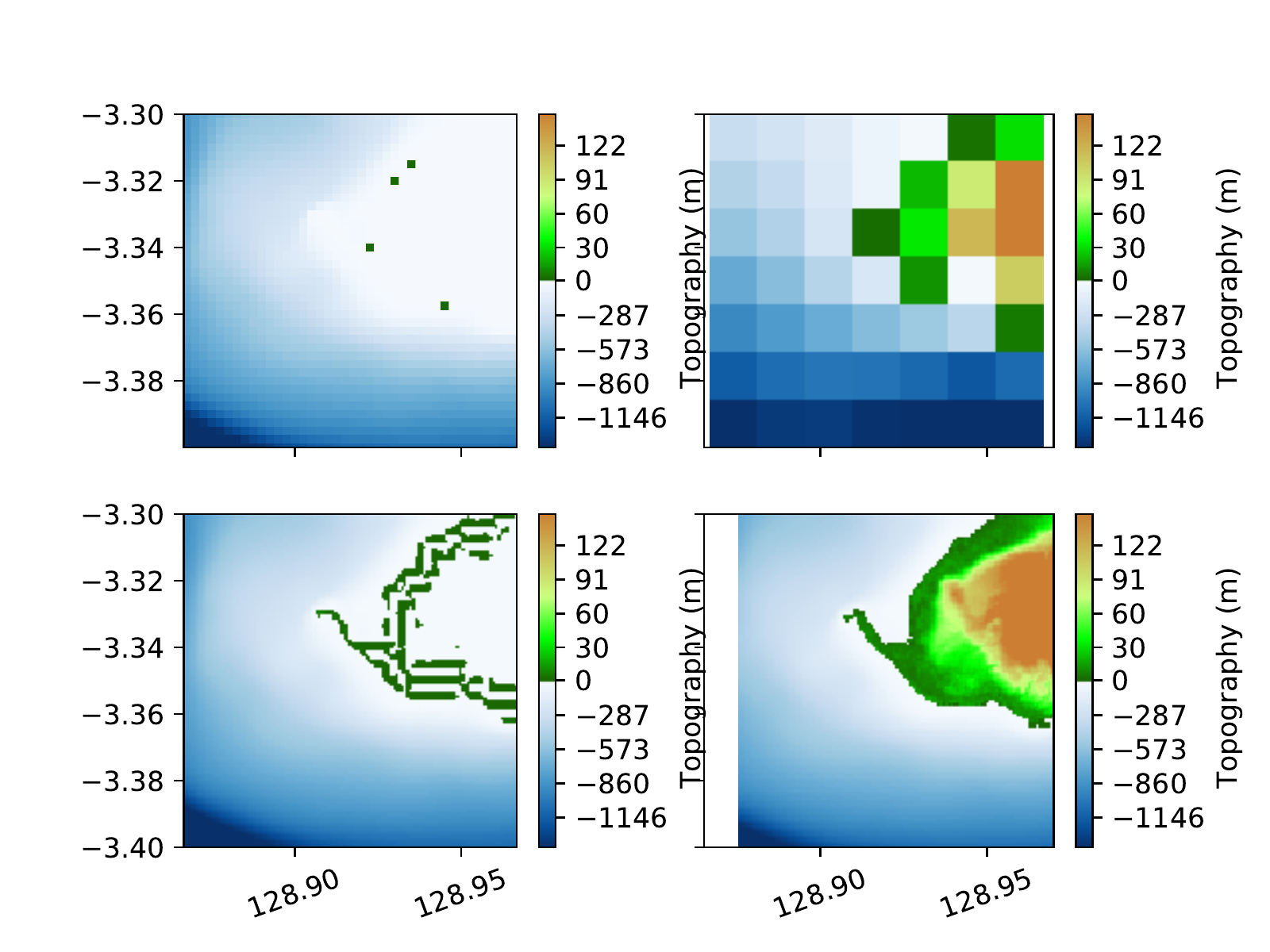}
  \caption{\label{fig:amahai_bathy} Combining all of the bathymetric and topographical sources into a single file for the bay near Amahai.  The upper left figure demonstrates the bathymetry drawn from the level curves exported from \url{http://inarisk.bnpb.go.id}.  The upper right figure shows the level of resolution for the NOAA bathymetry data.  The lower left figure shows the interpolation of these two data sets after zeroing out all positive values of the topography from the NOAA bathymetry.  Removing the positive topography creates some oscillatory behavior near sea level that is removed in the next step.  The lower right figure is the final product, combining the improved bathymetric data with the DEM coastline dataset. }
\end{figure}

In addition to these DEM coastline datasets and the NOAA bathymetry, we also took advantage of detailed sounding maps available from the Badan Nasional Penanggulangan Bencana (BNPB or Indonesian National Agency of Disaster Countermeasure, see \url{http://inarisk.bnpb.go.id}).  To convert this data into digitally accessible information, contours were taken from images exported from the website and then traced and interpolated in arcGIS to produce approximate depths in the same regions as 
the DEM files.  For example, the bathymetric readings based on this data are shown in Figure \ref{fig:amahai_bathy} for the bay of Amahai.  The upper left panel in Figure \ref{fig:amahai_bathy} depicts the bathymetry data that is gleaned from the BNPB and digitized by interpolating across contours of constant depth in arcGIS.  The upper right panel of Figure \ref{fig:amahai_bathy} depicts the bathymetry/topography from the NOAA bathymetry dataset.  Using the built in interpolative methods in Geoclaw's topotools package (\emph{topotools.interp\_unstructured} with the cubic interpolant, and a proximity radius of 1000), we interpolate the coastline and coarse bathymetry from the NOAA dataset to match the bathymetric contours from the upper right panel to produce the lower left panel.  This lower left panel does not accurately capture any of the topographical features of the coastline and suffers significantly from interpolant error onshore as there are no bathymetric readings there.  The actual shoreline and onshore topography is then overlaid from the DEM coastlines on top of the bottom left panel of Figure \ref{fig:amahai_bathy} to create the final product which is seen in the bottom right panel of the same Figure.  This retains the improved bathymetric contours, and yields an accurate coastline and near-shore topographical profile.

This same process is repeated for Palau Buru, and the coastline near the islands of Ambon, Saparua, Haruku, and Nasu Luat.  
All of these high resolution bathymetric files are used by Geoclaw when the wave approaches these locations onshore.


For the region near Banda Neira and Palau Ai, the bathymetric data was still quite rough, particularly for the narrow channels between Banda Neira, Banda Api, and Lonthor.  We obtained a set of soundings for this region from a map published by the Kepala Dinas Hidro-Oseanografi (the Indonesian Navy Hydrography and Oceanography Center) from data collected primarily in 1928/1929 \cite{banda_map2011}.   Using the same approach as described above for the bay of Amahai, these discrete soundings are interpolated for the entire region surrounding the Banda islands (except that a linear interpolant is used instead of cubic due to the sparsity of the measurements) and overlayed with the DEM coastlines.  


For the forward simulations of the tsunami wave, we employ an adjoint-based adaptive mesh strategy \cite{DavisLeVeque2016}.  This entails solving a linearized adjoint equation backward in time with sources centered at each gauge location  The solution of the adjoint equation produces waves that propagate backward in time from the desired observation locations to indicate what part of the forward wave will eventually influence the tsunami at those locations (see \cite{DavisLeVeque2016} for details).  To initialize the adjoint solver, we place a smoothed Gaussian perturbation $h(x,y)$ to the wave height at each gauge location given by:
\begin{equation}
    h(x,y) = \sum_k \exp(-r_k^2/150),
\end{equation}
where $r_k$ is the distance from the point $(x,y)$ to the gauge location $(x_k,y_k)$.  The solution of the linearized adjoint problem guides the choice of refinement regions of the fully nonlinear forward model, indicating where the wave that will reach the observed locations will be at specific times.  The benefit of using this approach as noted in \cite{DavisLeVeque2016} is that only those parts of the wave that will reach the desired locations are refined, i.e. the mesh refinement is restricted to those parts of the domain (in both space and time) that will most influence the final wave at the desired location.  In addition, for the application at hand, we only need to run the backward adjoint solver once, and then the generated output can be used for every sample so long as the gauge locations are not changed.  This saves a substantial amount of computational cost, allowing us to use a much finer mesh near the observational locations than a standard adaptive mesh would have allowed.

We use an adaptive mesh with 6 levels, starting with 6 arcminute resolution in the open water with no motion, and then going through $2\times,~ 2\times,~2\times,~ 3\times$, and $5\times$ grid refinements to those regions where the adjoint indicates the wave will be, resulting in the finest grid of 3 arcseconds which matches the fine resolution of the DEM coastline files.  This choice of refinement levels is made as it appears to be optimal computationally to reach 3 arcsecond resolution at the finest grid.  This means that the mesh levels are given by 6 arcminute, 3 arcminute, 1.5 arcminute, 45 arcsecond, 15 arcsecond, and 3 arcsecond resolution respectively.  In addition to this dynamic adaptation of the mesh, we statically fix regions near each gauge at the highest mesh resolution (3 arcseconds) for the entirety of the simulation, thus accurately capturing the wave characteristics near the observed locations.  
Implementation of such a highly refined grid for the region in question required some minor modification of the default list lengths in the fortran code as described in the code repository.  The backward adjoint solver is run on a 15 arcsecond grid and the output files are saved every 5 minutes to ensure adequate spatial and temporal resolution for the dynamic grid refinement.  Geoclaw interpolates these output files temporally to determine the wave location throughout the entire simulation.


All other settings in Geoclaw are set to their default
values.  An adaptive time step is adjusted
according to the Courant-Friedrichs-Lewy (CFL) condition with a desired CFL of $0.75$.  The spatial
discretization in Geoclaw is a second order scheme with the MC limiter
\cite{leveque2002finite} employed to avoid the development of
un-physical shocks.  All simulations are run for a physical time
window of $1.5$ hours to ensure that the wave has reached all of the
relevant locations (for this event the longest historically recorded time between
the earthquake and the arrival of the wave was approximately 40-45
minutes as shown in \cref{fig:likelihoods}).  Each simulation of
Geoclaw generates wave heights and arrival times at the locations
shown in \cref{fig:obslocations}.

\subsubsection{Wave inundation calculation} 
\label{sec:Inundation:model}
The observations of wave inundation at Banda Neira and Saparua are very precise, and seem to be important to infer the earthquake. Despite the precision of this measured distance, it is unclear what specific part of the shoreline these observations were recorded for, and even if this was clear, the highest resolution topography available isn't sufficient to completely trust a simulated wave inundation.  For these reasons, we opted to use a simplified model of wave inundation that is a deterministic function of the wave height on the shoreline.

The model that dictates the mapping from on-shore wave height to wave inundation length is taken from \cite{bryant2014} (equation
(2.15)), yielding a relationship between the maximum inundation
length, maximal wave height on-shore, and the average slope of the
shoreline.  This is given by:
\begin{align}
  \mathcal{I}_{\beta,n}(H) = \frac{k \cdot H^{1.33}\cos(\beta)}{n^2}
\label{eq:inundation} 
\end{align}
where $\mathcal{I}$ is the maximum landward inundation distance as a function of
the maximum shoreline wave height $H$.  Here $n$ is Manning's
coefficient of friction \cite{turcotte2002geodynamics} (selected using arcGIS and Google images near the
coastline in question, $\beta$ is the slope of the coastline and $k$
is an empirically determined constant equal to $0.06$.  We computed
$\beta$ as the average slope taken from a series of 1-dimensional
vertical profiles taken from arcGIS perpendicular to the coastline near
each gauge. This model was used to convert on-shore wave heights into on-shore inundation which is then evaluated against the likelihood probabilities of on-shore inundation from \cref{fig:likelihoods}.

\subsection{Sampling and Convergence} \label{sec:mcmc}
To quantify the results of the Bayesian inference, Markov Chain Monte Carlo (MCMC) was used to generate samples from the posterior measure. Because we did not have an adjoint solver for this PDE-based forward map, gradient-based methods like Hamiltonian Monte Carlo were not available. We therefore employed random walk-style Metropolis-Hastings MCMC with periodic Sequential Monte Carlo-style resampling according to posterior probability (see \cite[Section 5.3]{dashti2017bayesian}). This algorithm is summarized in \cref{alg:mcmc}. A diagonal covariance structure was used for the proposal kernel, with the step size in each of the six parameters tuned to approximate the optimal acceptance rate of roughly $0.23$ \cite[Section 12.2]{gelman2014bayesian}. The final standard deviations for the random walk proposal kernel are given in 
the GitHub repository.

\begin{algorithm}[H]
\caption{(MCMC as Applied to Tsunami Problem)}\label{alg:mcmc}
\begin{algorithmic}[1]
   \State Choose number of chains $M$, resampling rate $N$, proposal covariance $C$, and initial parameters $\unk_0^{(i)}, i=1,\dots,M$.
    \For{$k \geq 0$}
        \For{$i = 1,\dots,M$}
            \State Propose $\prop^{(i)} = \unk_k^{(i)} + \eta, \eta \sim N(0,C)$
            \State Run Geoclaw to compute likelihood $\llh(\fwd(\prop^{(i)}))$.
            \State Compute un-normalized posterior $\pst(\prop^{(i)})$ from \eqref{eq:post}.
            \State Set $\unk_{k+1}^{(i)} :=\prop^{(i)}$ with probability                $\min\{1,\pst(\prop^{(i)})/\pst(\unk_k^{(i)})\}$. 
            \State Otherwise take $\unk_{k+1}^{(i)} := \unk_{k}^{(i)}$.
        \EndFor
        \State If $k \mod N=0$, resample $\unk_k^{(i)} \sim \Sigma_j \pst(\unk_k^{(j)})\delta\left( \unk_k^{(j)} \right) /\Sigma_l \pst(\unk_k^{(l)}), i=1,\dots,M$.
        \State $k \to k + 1$.
    \EndFor
\end{algorithmic}
\end{algorithm}

To ensure that all viable seismic events were considered, we initialized fourteen (14) MCMC chains at locations around the Banda arc with initial magnitudes of either $8.5$ or $9.0$ Mw. Additional chains were tried at $8.0$ Mw; however, these were quickly discarded as they consistently failed to generate a sufficiently large wave to reach each of the observation points (\cref{fig:obslocations}) and therefore produced likelihoods of zero probability. Each of the 14 remaining chains was run for 24,000 samples, including a ``burn in'' of 6,000 samples, with resampling according to posterior probability every 6,000 samples. Our approximation to the posterior therefore is made up of a total of 252,000 samples. To ensure accurate statistics, chains were run well beyond when they appeared to have converged; for example, \emph{Gelman-Rubin diagnostic} $R$ \cite{gelman1992inference,gelman2014bayesian} for all parameters fell below 1.1 (a common convergence criterion) after 8,000 samples. Samples were computed using the compute resources available through BYU's Office of Research Computing, consuming a total of nearly 200,000 core-hours in all. 

\section{Results}
\label{sec:Results}

\subsection{Summary}


\begin{figure}[h]
    \centering
    \includegraphics[width=\textwidth]{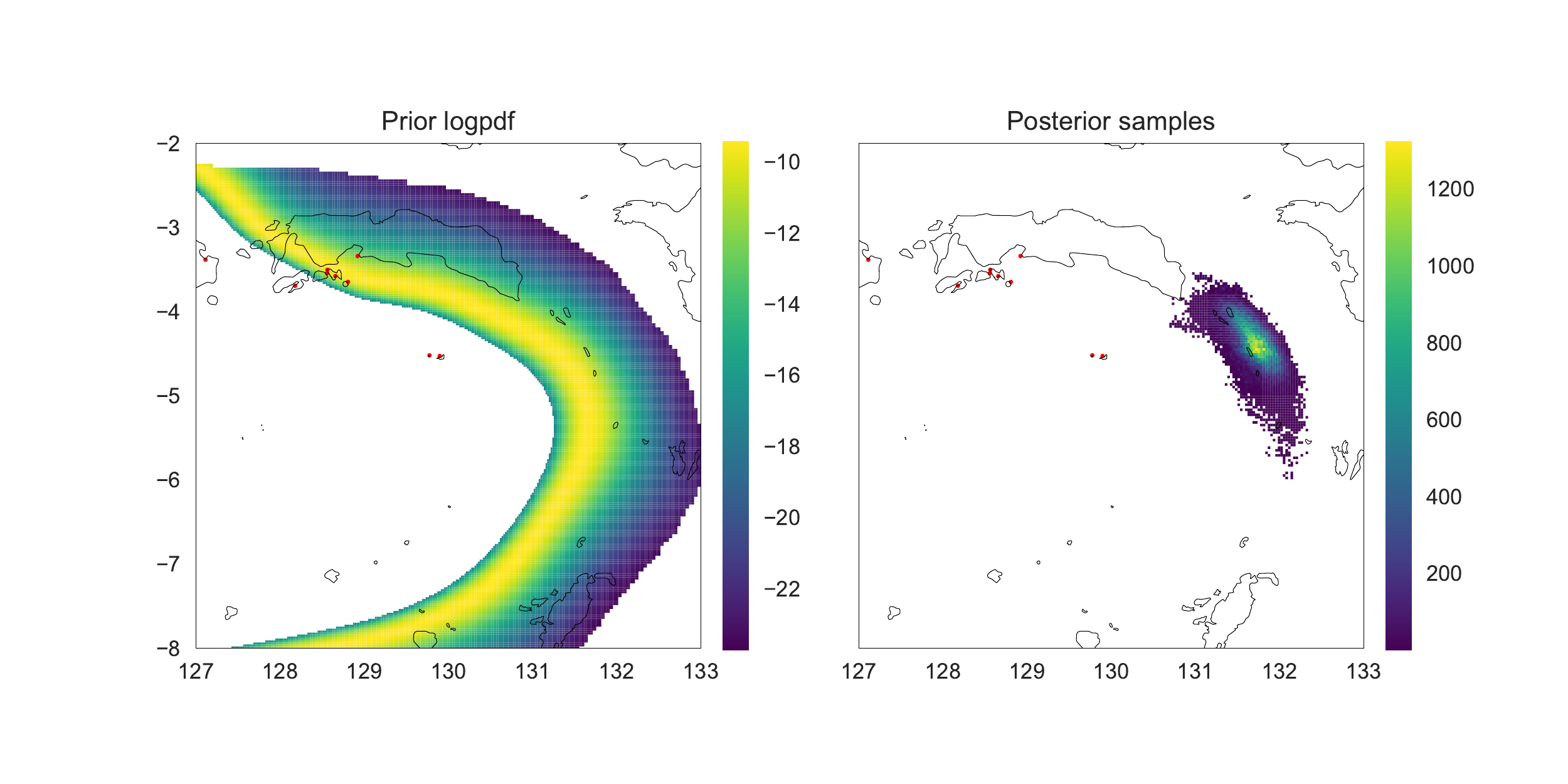}
    \caption[Latitude/longitude prior and posterior]{Sample centroids from the posterior compared with the prior in latitude/longitude (the same region in the green rectangle of \cref{fig:obslocations}). The posterior is concentrated in a small region in the northeast of the arc.  Note that the color gradient is not the same quantitatively between these two plots, but the general characterization is accurate, i.e. the prior is distributed evenly over the entire arc (via the depth calculation) while the posterior is heavily concentrated in the northeastern section of the arc.}
    \label{fig:latlonpost}
\end{figure}

\begin{figure}[h]
    \centering
    \includegraphics[width=.7\textwidth]{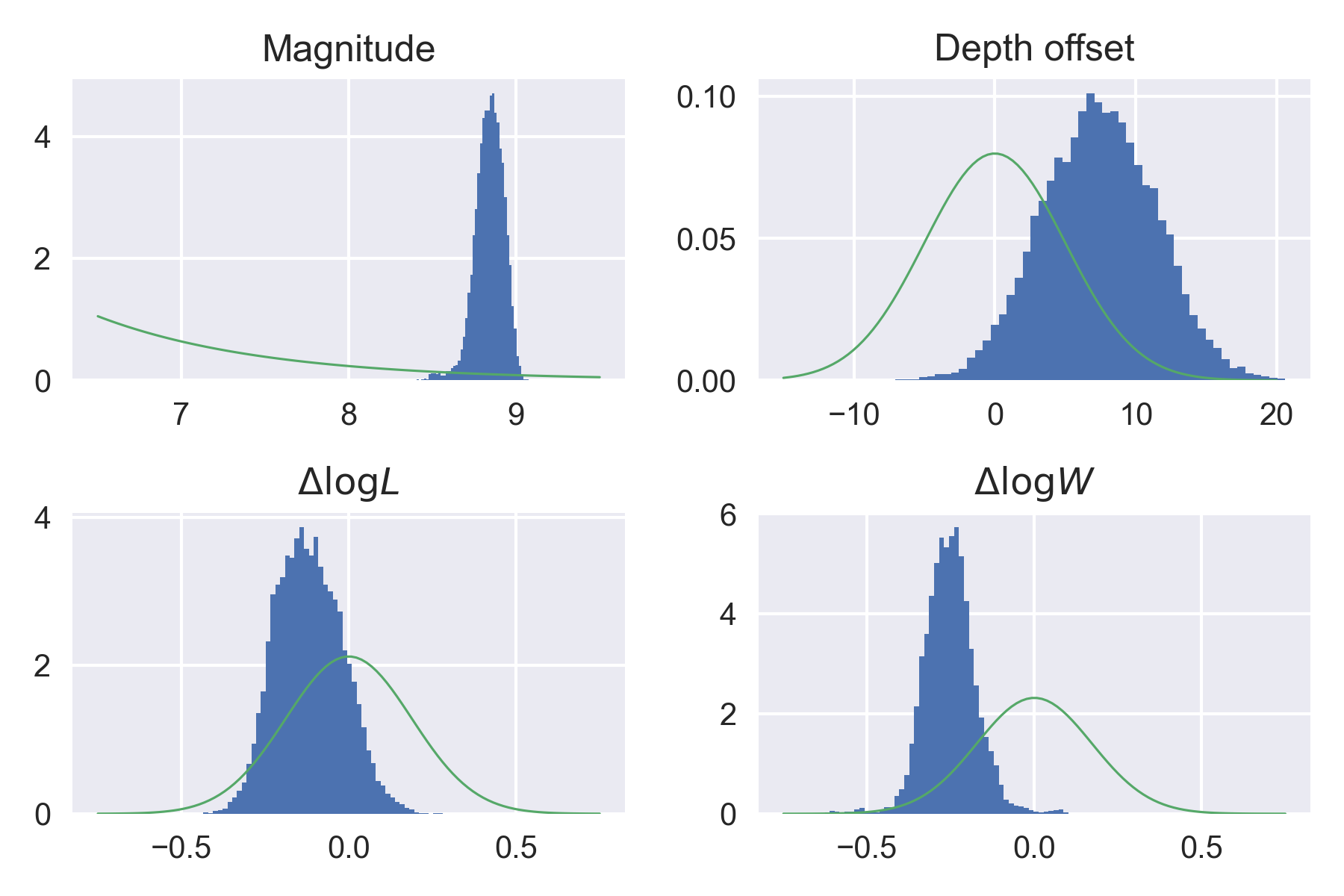}
    \caption[Priors and posteriors for other parameters]{Magnitude, depth offset, $\Delta \log L$, and $\Delta \log W$ posterior histograms, compared to the associated prior distribution densities (plotted in green)  The bottom two plots are histograms of the slip and depth (in meters for both) drawn from the posterior.}
    \label{fig:priorvspost}
\end{figure}

The results of the inference are summarized in Figures \ref{fig:latlonpost} and \ref{fig:priorvspost} and show stark differences between the prior and posterior distributions. Whereas the prior encompassed all parts of the Banda arc with a reasonable depth, the posterior for the centroid is narrowly concentrated in a region near 4.5\degree S, 131.5\degree E, which is situated in a shallow part of the subduction interface. That is, the modeling implies that the centroid must have been located within this region to best match the observations summarized in \cref{fig:likelihoods}. Also notable is the marginal posterior for magnitude: despite a prior that heavily preferred lower magnitudes, the posterior is concentrated around earthquakes of magnitude 8.8. This is because the tsunami simulations indicated that a large event was required to produce the wave heights described in the historical accounts at several of the observation locations.

More subtle inference is seen in magnitude-normalized length and width. The posterior favors rupture zones that are relatively narrow for their magnitude, i.e., the length and width are smaller than is typical, given the magnitude of the event. This is likely because the inference is trying to balance observations of wave height and arrival time. Simulations indicate that an earthquake needs to be quite large in order to produce the observed wave heights in, for instance, Banda Neira. However, larger earthquakes, all else being equal, have rupture zones that are closer to Banda Neira, thus reducing the arrival time of the wave. The posterior therefore favors a smallish rupture zone given the magnitude (keeping the arrival time in check) balanced by a larger slip to achieve the observed wave heights.

\subsection{Fault Characteristics by Centroid Location}
\begin{figure}[h]
    \centering
    \includegraphics[width=\textwidth]{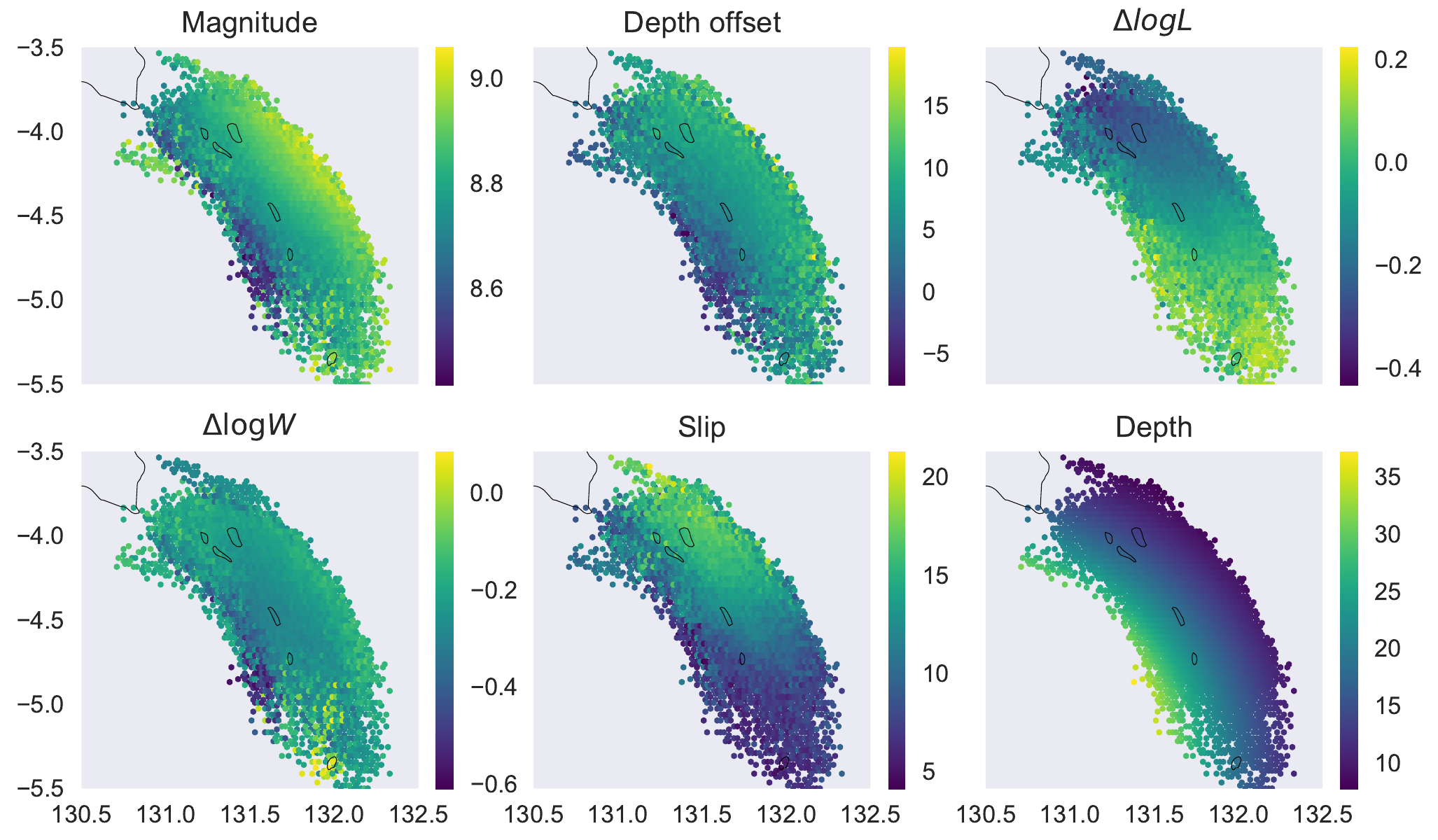}
    \caption[Sample parameter conditional expectations against latitude and longitude]{Posterior conditional expectation of sample parameters magnitude (Mw), depth offset (km), $\Delta \log L$, and $\Delta \log W$ as well as the slip (m) and depth (km) conditioned on latitude/longitude, shown for the same region as the yellow rectangle in \cref{fig:obslocations}.}
    \label{fig:samp_cond}
\end{figure}
We now describe how the fault parameters change with geographic location according to the computed posterior. \cref{fig:samp_cond} displays the approximate conditional expectation for the sample parameters magnitude, depth offset, $\Delta \log L$, and $\Delta \log W$, conditioned on latitude and longitude as well as the conditional expectation for the model parameters slip and depth; these figures therefore show the expected value of each parameter if we were to assume a given centroid location. Several trends are apparent:
\begin{itemize}
\item \textbf{The farther outside the arc, the higher the expected magnitude.} This is not surprising, as higher magnitudes would be required to produce large enough waves at that distance from the observation points. 
\item \textbf{The farther outside the arc, the greater the value of depth offset.} This appears to counteract the shallowing of the fault towards the outside of the arc, ultimately producing earthquakes at nearly constant depth among accepted samples.  This can be seen even more clearly in the conditional expectation of depth, where there are variations in the depth, but they are relatively mild (most of the points in this plot are on the shallower end of the colorbar).
\item \textbf{The closer the center of the rupture is to the coast of Seram, the shorter the length (smaller $\Delta \log L$), and higher the slip of the rupture.} This is likely due to the rupture extending underneath Seram Island, which leads to a smaller tsunami (as only some of the rupture occurs beneath the ocean). Thus, a shorter rupture zone increases the slip as seen in the conditional expectation of slip (and thus wave height), counteracting the influence of Seram Island on the tsunami generation.  On the other hand, the earthquakes to the south or west have lower slip values as the tsunami source is then closer to the observation locations.
\end{itemize}

\subsection{Forward Model and Output}

\begin{figure}[ht]
    \centering
    \includegraphics[width=\textwidth]{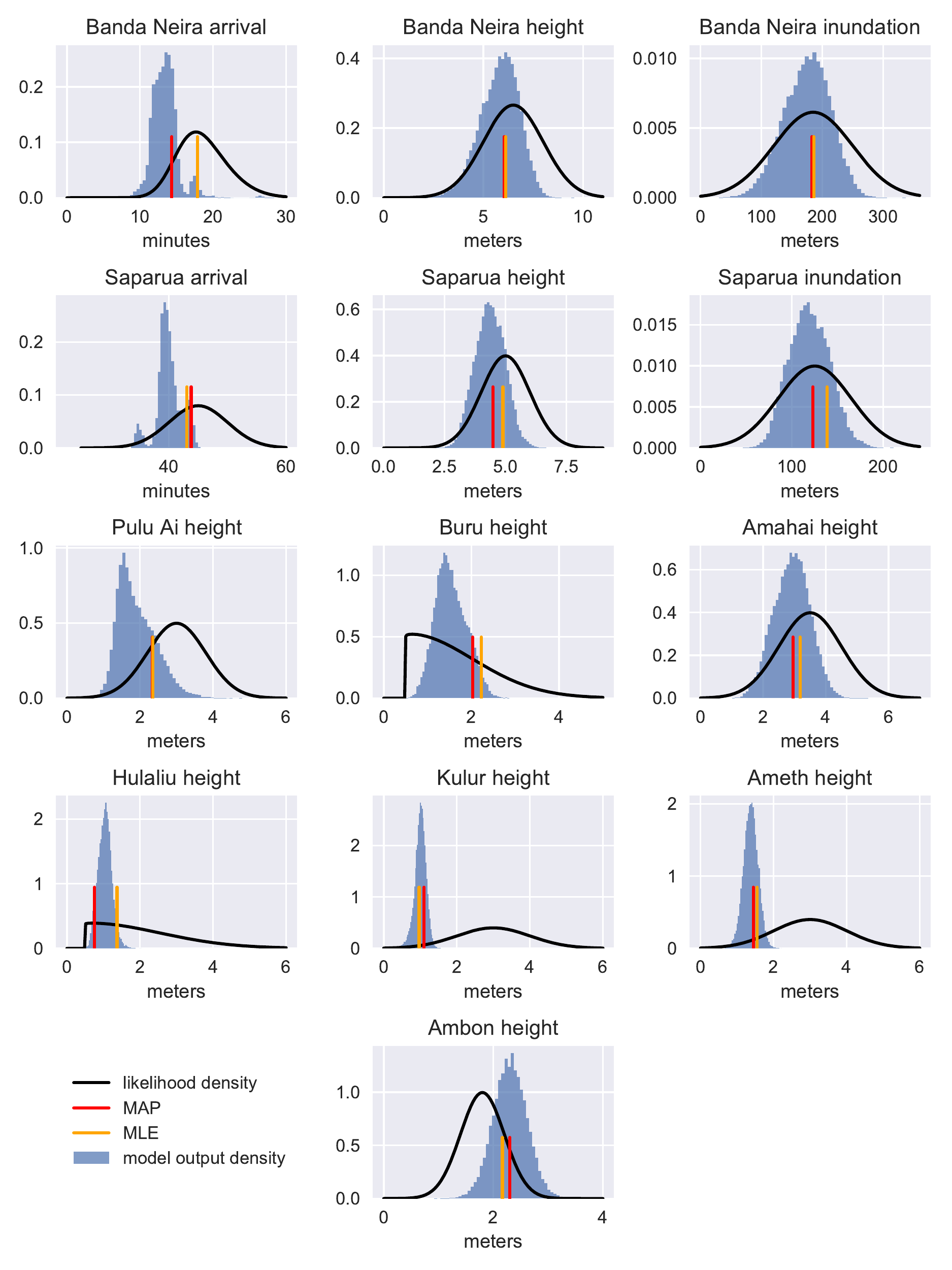}
    \caption[Model output compared to likelihood densities]{Model output compared to likelihood densities. The blue histograms represent the forward model outputs corresponding to the posterior distribution. The black curves are the likelihood densities assigned to each observation. The model outputs corresponding to the estimated maximum a posteriori (MAP) point and maximum likelihood estimate (MLE) are marked with red and orange lines, respectively.}
    \label{fig:obs_hists}
\end{figure}

It is also worth considering the implied observation distributions, known as the \emph{posterior predictive distributions}. This has two primary interpretations, first in understanding the drivers of the results, and second as an assessment of hazard risk. We describe each of these in turn. 

\textbf{Understanding the Results.} Comparing the posterior predictive distributions and the likelihood distributions, shown in \cref{fig:obs_hists}, helps describe what drove the model's conclusions. Banda Neira and Saparua provided the largest contribution to the likelihood, and we see that the posterior samples broadly matched our interpretation of the observations there. The posterior samples at Kulur and Ameth stand out as different from the proposed likelihood, with waves smaller than our interpretation of the accounts. However this is acceptable and actually validates the current approach, given that these accounts were not specific, and we assigned wide distributions to them, indicating that the posterior distribution matched the likelihood from the other locations but was unable to match the proposed distributions at these locations. Overall, the posterior distribution is consistent with the observations recorded in our sources, with sufficient differences to make the inferred posterior distribution distinct from both the likelihood density and the prior distribution.  This indicates that the posterior is not just a re-sampling of the likelihood and/or prior but has instead found earthquakes that best match the data and physical limitations of the situation.  

Additional insights into the limitations of our approach can be gleaned from the differences between the posterior predictive distribution and the likelihood.  For instance in addition to the differences for the wave height at certain locations, we notice a significant difference between the distributions for arrival time at both Banda Neira and Saparua.  In fact, it appears that our assigned likelihood distribution over estimated the actual arrival time.  Such a discrepancy is not unexpected however, because the current forward model assumes an instantaneous rupture whereas the historical account implies that the earthquake lasted approximately 5 minutes which would adequately account for the shifted distributions shown in \cref{fig:obs_hists}.

\textbf{Hazard Risk.} The posterior predictive distribution can also be interpreted as an estimate of the hazard risk for the specified observation locations.  The histograms in \cref{fig:obs_hists} represent what communities in these locations might be expected to experience -- the wave heights, arrival times, and inundations -- should a similar event happen in the future.  For instance, if an event of this magnitude occurred in the same location on the Banda arc, we anticipate a wave of approximately $2.5m$ to reach the populous city of Ambon (approximately 300,000 people).  For those living in the bay of Ambon, this gives a probabilistic hazard assessment that can be coupled with detailed topographical information to assess potential flood levels as well as economic and societal impact from a similar future event.

\subsection{Claim \& Corroborating Evidence}

The implied claim of our posterior distribution is this: if the 1852 Banda arc tsunami was caused by a mega-thrust subduction zone earthquake, it was a magnitude $\sim$8.8 mega-thrust event centered near 4.5\degree S, 131.5\degree E, and this type of event matches the historical account quite well. During the analysis, we discovered an item of corroborating evidence for this claim, in the form of the Slab2 depth uncertainty data. The Slab2 model of the subduction zone is based on seismically collected instrumental data that can be used to infer the interface geometry. The more earthquakes that have occurred recently on a particular segment of a fault, the more certain we can be of the geometry. Regions of uncertainty correspond to ``seismic gaps''; fault segments that have been relatively silent during the modern period of instrumental data collection. A seismic gap may represent a location where hundreds of years of stress has accumulated, which eventually results in a large earthquake when the fault slips and the stress is released \cite{seismic-gap}. While not all seismic gaps turn out to be dangerous \cite{kagan1991}, they are still important to consider as possible sources for an event such as the 1852 Banda arc earthquake.

\begin{figure}
    \centering
    \includegraphics[width=\textwidth]{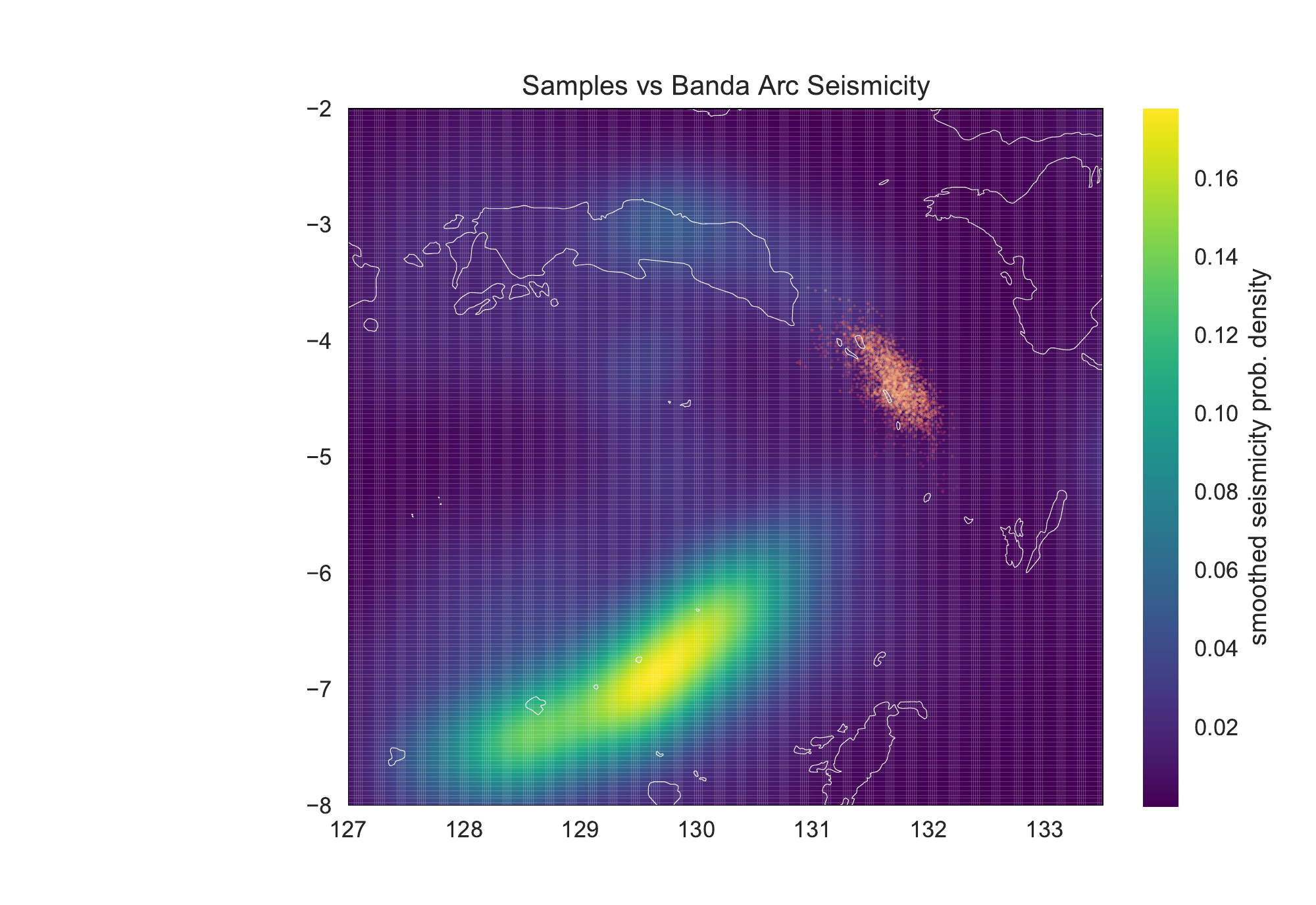}
    \caption{Banda arc Slab2 seismicity compared to the location of samples from the posterior distribution.  The seismicity is computed using a Gaussian kernel density estimator from locations of instrumentally recorded earthquakes (as provided in the Slab2 dataset) and is displayed on a blue-yellow color gradient with the dark blue regions referring to areas where little to no seismic events were recorded, and yellow being the highest concentration of seismic happenings as indicated in the colorbar.  The logarithm of the expected posterior distribution conditioned on the latitude-longitude location is shown in the orange color scheme.  Clearly the posterior distribution is concentrated in a region of low instrumentally recorded seismicity.  This is the same region as the green rectangle in \cref{fig:obslocations}.}
    \label{fig:seismic-gap}
\end{figure}

Both the Slab2 depth uncertainty, and the underlying seismic dataset, demonstrate the presence of a seismic gap in the region where our posterior distribution is concentrated (see \cref{fig:seismic-gap}). This can be viewed as evidence, distinct and separate from our usage of Slab2, that supports the results of our analysis.

\section{Discussion and Future Work}\label{sec:discussion}
This paper has presented a systematic approach to determining the strength and location of historically observed mega-thrust earthquakes via observations of the resultant tsunami.  The Bayesian nature of this investigation not only provides an understanding of the earthquake parameters that may have caused this event, but also yields information regarding the uncertainty of these post-dicted parameters, and the correlations between them.  A significant attribute of this approach is that the necessary assumptions are incorporated directly into the method and are explicitly described.  We note that the 1852 Banda arc event is certainly not the only event for which this methodology is feasible.  There are several  other events both in Indonesia and elsewhere for which there is either historical or geological evidence for significant seismic events, but little or no instrumental data to draw from.  In addition, although this article focuses on mega-thrust events, the methodology is sufficiently flexible to work for a submarine slump induced tsunami, or any other type of hazard for which a forward model is available.  We will look into these possibilities in future studies, including analyzing the potential for the 1852 event to have been caused by a submarine slump as suggested by \cite{Cummins2020,PranantyoCummins2020}.

One of the goals of applying a Bayesian approach to this inverse problem was to regularize it -- given the anecdotal nature of the data, it was important to minimize the impact that small changes in the assumptions would have on the resulting posterior measure. This concept is known as ``robustness'' of Bayesian inference. Some theoretical results -- see, e.g., \cite[Chapter 4]{stuart2010inverse} -- show that small changes in the assumptions provably result in small changes in the posterior measure. We therefore expect that the results presented in this paper will exhibit this kind of well-behaved local sensitivity. However, Bayesian robustness is in itself a line of research -- see, for example, \cite{berger1994overview,insua2000robust} -- and there are some cases where Bayesian inference can be ``brittle'' \cite{owhadi2015brittleness}. We intend, therefore, to conduct a computational study of local sensitivity to assumptions, and in particular to changes in the prior and likelihood distributions. However, given the computational cost of computing one posterior, effectively computing several posterior measures will require a number of approximations to speed up the forward map and MCMC convergence. We are currently working on those steps and plan to present the results of that study in a future paper.

We note that since the 1852 earthquake investigated here, the Banda arc region has seen an order of magnitude increase in population, increasing the seismic and tsunami disaster potential for the inhabitants.  However, risk assessments based on the short (geologically speaking) period of instrumental records for this and other densely populated regions (i.e. Java and Bali) underestimate the seismic and tsunami disaster potential.  Some of these regions have not experienced mega-thrust earthquakes for several hundred years.  At the current rates of seismic loading on these subduction zones, enough elastic strain has accumulated to cause another Mw 8.5-9.0 event akin to that described here in the Banda Sea.  This fact, coupled with the evidence provided here, indicate that at least in the case of the Banda arc, mega-thrust events are not only possible, but highly likely to have occurred in the past and thus likely to recur in the future.

\FloatBarrier

\section*{Acknowledgements}
Datasets for this research are available in these in-text data citation references: \cite{zenodov1}.

The authors acknowledge Advanced Research Computing
at Virginia Tech (\url{http://www.arc.vt.edu}) and Research Computing at BYU (\url{http://rc.byu.edu}) for providing
computational resources and technical support that have contributed to
the results reported within this paper.  JPW and RH would like to
thank the Office or Research and Creative Activities at BYU for
supporting several of the students' efforts on this project through a
Mentoring Environment Grant, as well as generous support from the
College of Physical and Mathematical Sciences and the Mathematics and Geology Departments. We also acknowledge the
visionary support of Geoscientists Without Borders.  JPW would like to thank
Tulane University for supporting him as a visiting scholar during the 
Spring 2018 semester where a portion of this work was carried out.  JPW was partially supported by the Simons Foundation travel grant under 586788.  NEGH was also partially supported by NSF Grants DMS-1313272, DMS-1816551, and the Simons Foundation travel grant under 515990.

We would like to thank our colleagues W. F. Christensen, J. Borggaard,
E. J. Evans, R. LeVeque, V. Martinez, S. McKinley, C. Mondaini,
C. S. Reese, G. Simpson \& F. Viens for valuable feedback on this
work.  In addition to the authors, there were several students at BYU that provided feedback and assistance to various parts of this project including: A. (Oveson) Bagley, R. Howell, J. Lapicola, C. Carter, I. Sorenson, B. Berrett, C. Ringer, J. Voorhees, and Y. Zhao.
 We also acknowledge the feedback of two referees whose suggestions greatly improved the presentation of the results.

\bibliographystyle{plain}
\bibliography{tsunami}

\begin{thebibliography}{10}

\bibitem{banda_map2011}
Detailed soundings of the banda islands, 2011.
\newblock Based on data collected by Dinaa Hidrografi Negara Belanda (The
  Hydrographic Service of the Royal Netherlands Navy) in 1928/1929.

\bibitem{NCEI_tsunami}
{Tsunami Sources 1610 B.C. to A.D. 2017 from Earthquakes, Volcanic Eruptions,
  Landslides, and Other Causes}.
\newblock
  \url{https://www.ngdc.noaa.gov/hazard/data/publications/tsunami-sources-2017.pdf},
  2017.
\newblock Accessed: 2020-09-01.

\bibitem{anon1940}
Anonymous.
\newblock {Aardbevingen in den Oost Indischen Archipel waargenomen gedurende
  het jaar 1938}.
\newblock {\em Natuurk. Tijdschr. Ned. Indie}, 40:38--74, 1940.

\bibitem{USGSmaggroup}
William Bakun.
\newblock {\em {USGS Earthquake Magnitude Working Group}}, 2002 (accessed July
  19, 2020).

\bibitem{barkan2010tsunami}
Roy Barkan and Uri Ten~Brink.
\newblock Tsunami simulations of the 1867 virgin island earthquake: Constraints
  on epicenter location and fault parameterstsunami simulations of the 1867
  virgin island earthquake: Constraints on epicenter location.
\newblock {\em Bulletin of the Seismological Society of America},
  100(3):995--1009, 2010.

\bibitem{berger1994overview}
James~O Berger, El{\'\i}as Moreno, Luis~Raul Pericchi, M~Jes{\'u}s Bayarri,
  Jos{\'e}~M Bernardo, Juan~A Cano, Juli{\'a}n De~la Horra, Jacinto
  Mart{\'\i}n, David R{\'\i}os-Ins{\'u}a, Bruno Betr{\`o}, et~al.
\newblock An overview of robust {B}ayesian analysis.
\newblock {\em Test}, 3(1):5--124, 1994.

\bibitem{berger2011geoclaw}
Marsha~J Berger, David~L George, Randall~J LeVeque, and Kyle~T Mandli.
\newblock {The GeoClaw software for depth-averaged flows with adaptive
  refinement}.
\newblock {\em Advances in Water Resources}, 34(9):1195--1206, 2011.

\bibitem{beskos2009optimal}
Alexandros Beskos, Gareth Roberts, and Andrew Stuart.
\newblock {Optimal scalings for local Metropolis--Hastings chains on nonproduct
  targets in high dimensions}.
\newblock {\em The Annals of Applied Probability}, 19(3):863--898, 2009.

\bibitem{bock2003crustal}
Yehuda Bock, L~Prawirodirdjo, J~F Genrich, C~W Stevens, R~McCaffrey, C~Subarya,
  S~S~O Puntodewo, and E~Calais.
\newblock {Crustal motion in Indonesia from global positioning system
  measurements}.
\newblock {\em Journal of Geophysical Research: Solid Earth}, 108(B8), 2003.

\bibitem{bondevik2008earth}
Stein Bondevik.
\newblock {Earth science: The sands of tsunami time}.
\newblock {\em Nature}, 455(7217):1183, 2008.

\bibitem{bryant2007cosmogenic}
E~Bryant, Grant Walsh, and Dallas Abbott.
\newblock {Cosmogenic mega-tsunami in the Australia region: are they supported
  by Aboriginal and Maori legends?}
\newblock {\em Geological Society, London, Special Publications},
  273(1):203--214, 2007.

\bibitem{bryant2014}
Edward Bryant.
\newblock {\em {Tsunami: the underrated hazard}}.
\newblock Springer, 2014.

\bibitem{carter1976stratigraphical}
David~J Carter, Michael~G Audley-Charles, and AJ~Barber.
\newblock {Stratigraphical analysis of island arc—continental margin
  collision in eastern Indonesia}.
\newblock {\em Journal of the Geological Society}, 132(2):179--198, 1976.

\bibitem{Cummins2020}
Phil~R Cummins, Ignatius~R Pranantyo, Jonathan~M Pownall, Jonathan~D Griffin,
  Irwan Meilano, and Siyuan Zhao.
\newblock Earthquakes and tsunamis caused by low-angle normal faulting in the
  banda sea, indonesia.
\newblock {\em Nature Geoscience}, 13(4):312--318, 2020.

\bibitem{dashti2017bayesian}
Masoumeh Dashti and Andrew~M Stuart.
\newblock {The {B}ayesian approach to inverse problems}.
\newblock {\em Handbook of Uncertainty Quantification}, pages 311--428, 2017.

\bibitem{DavisLeVeque2016}
B~Davis and R~LeVeque.
\newblock Adjoint methods for guiding adaptive mesh refinement in tsunami
  modeling.
\newblock {\em Pure \& Applied Geophysics}, 173(12), 2016.

\bibitem{fisherharris2016}
TszMan~L Fisher and Ron~A Harris.
\newblock {Reconstruction of 1852 Banda Arc megathrust earthquake and tsunami}.
\newblock {\em Natural Hazards}, 83(1):667--689, 2016.

\bibitem{fujii2007tsunami}
Yushiro Fujii and Kenji Satake.
\newblock Tsunami source of the 2004 sumatra--andaman earthquake inferred from
  tide gauge and satellite data.
\newblock {\em Bulletin of the Seismological Society of America},
  97(1A):S192--S207, 2007.

\bibitem{fukuda2008fully}
Jun’ichi Fukuda and Kaj~M Johnson.
\newblock A fully bayesian inversion for spatial distribution of fault slip
  with objective smoothing.
\newblock {\em Bulletin of the Seismological Society of America},
  98(3):1128--1146, 2008.

\bibitem{gelman2014bayesian}
Andrew Gelman, John~B Carlin, Hal~S Stern, and Donald~B Rubin.
\newblock {\em {Bayesian data analysis}}, volume~2.
\newblock Taylor \& Francis, 2014.

\bibitem{gelman1992inference}
Andrew Gelman, Donald~B Rubin, et~al.
\newblock Inference from iterative simulation using multiple sequences.
\newblock {\em Statistical Science}, 7(4):457--472, 1992.

\bibitem{giraldi2017bayesian}
Lo{\"\i}c Giraldi, Olivier~P Le~Ma{\^\i}tre, Kyle~T Mandli, Clint~N Dawson,
  Ibrahim Hoteit, and Omar~M Knio.
\newblock Bayesian inference of earthquake parameters from buoy data using a
  polynomial chaos-based surrogate.
\newblock {\em Computational Geosciences}, 21(4):683--699, 2017.

\bibitem{gonzalez2011validation}
Frank~I Gonz{\'a}lez, Randall~J LeVeque, Paul Chamberlain, Bryant Hirai,
  Jonathan Varkovitzky, and David~L George.
\newblock {Validation of the geoclaw model}.
\newblock In {\em {NTHMP MMS Tsunami Inundation Model Validation Workshop.
  GeoClaw Tsunami Modeling Group}}, 2011.

\bibitem{GrNgCuCi2018}
Jonathan Griffin, Ngoc Nguyen, Phil Cummins, and Athanasius Cipta.
\newblock {Historical Earthquakes of the Eastern {S}unda Arc: Source Mechanisms
  and Intensity-Based Testing of {I}ndonesia's National Seismic Hazard
  Assessment}.
\newblock {\em Bulletin of the Seismological Society of America}, 109, 12 2018.

\bibitem{grimes2006mapping}
Barbara~Dix Grimes.
\newblock {. Mapping Buru: The Politics of Territory and Settlement on an
  Eastern Indonesian Island}.
\newblock {\em Sharing the earth, dividing the land: land and territory in the
  Austronesian world}, pages 135--155, 2006.

\bibitem{hamilton1979tectonics}
Warren~Bell Hamilton.
\newblock {\em Tectonics of the Indonesian region}, volume 1078.
\newblock US Government Printing Office, 1979.

\bibitem{hamzah2000tsunami}
Latief Hamzah, Nanang~T Puspito, and Fumihiko Imamura.
\newblock Tsunami catalog and zones in indonesia.
\newblock {\em Journal of Natural Disaster Science}, 22(1):25--43, 2000.

\bibitem{kanamori1979}
Thomas~C. Hanks and Hiroo Kanamori.
\newblock A moment magnitude scale.
\newblock {\em Journal of Geophysical Research: Solid Earth},
  84(B5):2348--2350, 1979.

\bibitem{harris1991temporal}
RA~Harris.
\newblock {Temporal distribution of strain in the active Banda orogen: a
  reconciliation of rival hypotheses}.
\newblock {\em Journal of Southeast Asian Earth Sciences}, 6(3-4):373--386,
  1991.

\bibitem{harris2011nature}
Ron Harris.
\newblock {The nature of the Banda Arc--continent collision in the Timor
  region}.
\newblock In {\em Arc-Continent Collision}, pages 163--211. Springer, 2011.

\bibitem{harris2016waves}
Ron Harris and Jonathan Major.
\newblock {Waves of destruction in the East Indies: the Wichmann catalogue of
  earthquakes and tsunami in the Indonesian region from 1538 to 1877}.
\newblock {\em Geological Society, London, Special Publications}, 441:SP441--2,
  2016.

\bibitem{HayesSlab2}
Gavin~P. Hayes, Ginevra~L. Moore, Daniel~E. Portner, Mike Hearne, Hanna Flamme,
  Maria Furtney, and Gregory~M. Smoczyk.
\newblock Slab2, a comprehensive subduction zone geometry model.
\newblock {\em Science}, 362(6410):58--61, 2018.
\newblock Slab2 GitHub: \url{https://github.com/usgs/slab2}, data download:
  \url{https://www.sciencebase.gov/catalog/item/5aa1b00ee4b0b1c392e86467}.

\bibitem{heuret2012relation}
Arnauld Heuret, CP~Conrad, F~Funiciello, Serge Lallemand, and L~Sandri.
\newblock Relation between subduction megathrust earthquakes, trench sediment
  thickness and upper plate strain.
\newblock {\em Geophysical Research Letters}, 39(5), 2012.

\bibitem{heuret2011physical}
Arnauld Heuret, Serge Lallemand, Francesca Funiciello, Claudia Piromallo, and
  Claudio Faccenna.
\newblock {Physical characteristics of subduction interface type seismogenic
  zones revisited}.
\newblock {\em Geochemistry, Geophysics, Geosystems}, 12(1), 2011.

\bibitem{hilton1989helium}
DR~Hilton and H~Craig.
\newblock {A helium isotope transect along the Indonesian archipelago}.
\newblock {\em Nature}, 342(6252):906--908, 1989.

\bibitem{insua2000robust}
David~R{\'\i}os Insua and Fabrizio Ruggeri.
\newblock {\em Robust Bayesian Analysis}, volume 152.
\newblock Springer Science \& Business Media, 2000.

\bibitem{jankaew2008medieval}
Kruawun Jankaew, Brian~F Atwater, Yuki Sawai, Montri Choowong, Thasinee
  Charoentitirat, Maria~E Martin, and Amy Prendergast.
\newblock {Medieval forewarning of the 2004 Indian Ocean tsunami in Thailand}.
\newblock {\em Nature}, 455(7217):1228, 2008.

\bibitem{Kagan2002}
Yan~Y. Kagan.
\newblock {Seismic moment distribution revisited: I. Statistical results}.
\newblock {\em Geophysical Journal International}, 148(3):520--541, 03 2002.

\bibitem{kagan1991}
Yan~Y. Kagan and David~D. Jackson.
\newblock Seismic gap hypothesis: Ten years after.
\newblock {\em Journal of Geophysical Research: Solid Earth},
  96(B13):21419--21431, 1991.

\bibitem{kaipio2005statistical}
Jari Kaipio and Erkki Somersalo.
\newblock {\em {Statistical and computational inverse problems}}, volume 160 of
  {\em {Applied Mathematical Sciences}}.
\newblock Springer Science \& Business Media, 2005.

\bibitem{kubota2018tsunami}
Tatsuya Kubota, Wataru Suzuki, Takeshi Nakamura, Naotaka~Y Chikasada, Shin Aoi,
  Narumi Takahashi, and Ryota Hino.
\newblock Tsunami source inversion using time-derivative waveform of offshore
  pressure records to reduce effects of non-tsunami components.
\newblock {\em Geophysical Journal International}, 215(2):1200--1214, 2018.

\bibitem{leveque2002finite}
Randall~J LeVeque.
\newblock {\em {Finite volume methods for hyperbolic problems}}, volume~31.
\newblock Cambridge university press, 2002.

\bibitem{leveque2008high}
Randall~J LeVeque and David~L George.
\newblock {High-resolution finite volume methods for the shallow water
  equations with bathymetry and dry states}.
\newblock In {\em {Advanced numerical models for simulating tsunami waves and
  runup}}, pages 43--73. World Scientific, 2008.

\bibitem{leveque2011tsunami}
Randall~J. LeVeque, David~L. George, and Marsha~J. Berger.
\newblock {Tsunami modelling with adaptively refined finite volume methods}.
\newblock {\em Acta Numerica}, 20:211--289, 2011.

\bibitem{liu2008monte}
Jun~S Liu.
\newblock {\em {Monte Carlo strategies in scientific computing}}.
\newblock Springer Science \& Business Media, 2008.

\bibitem{LiuHarris2014}
Z.~Y.~C. Liu and R.~A. Harris.
\newblock {Discovery of possible mega-thrust earthquake along the Seram Trough
  from records of 1629 tsunami in eastern Indonesian region}.
\newblock {\em Natural Hazards}, 72:1311--1328, 2014.

\bibitem{malinverno2002parsimonious}
Alberto Malinverno.
\newblock Parsimonious bayesian markov chain monte carlo inversion in a
  nonlinear geophysical problem.
\newblock {\em Geophysical Journal International}, 151(3):675--688, 2002.

\bibitem{martin2019reassessment}
Stacey~Servito Martin, Linlin Li, Emile~A. Okal, Julie Morin, Alexander E.~G.
  Tetteroo, Adam~D. Switzer, and Kerry~E. Sieh.
\newblock {Reassessment of the 1907 Sumatra "Tsunami Earthquake" Based on
  Macroseismic, Seismological, and Tsunami Observations, and Modeling}.
\newblock {\em Pure and Applied Geophysics}, 2019.

\bibitem{mccaffrey2007next}
Robert McCaffrey.
\newblock The next great earthquake.
\newblock {\em SCIENCE-NEW YORK THEN WASHINGTON-}, 315(5819):1675, 2007.

\bibitem{seismic-gap}
W.~R. McCann, S.~P. Nishenko, L.~R. Sykes, and J.~Krause.
\newblock Seismic gaps and plate tectonics: Seismic potential for major
  boundaries.
\newblock {\em {Pure and Applied Geophysics}}, 117(6):1082--1147, 1979.

\bibitem{Mc1999}
K.~McCue.
\newblock {Seismic hazard mapping in Australia, the Southwest Pacific and
  Southeast Asia}.
\newblock {\em Annals of Geophysics}, 42:1191--1198, 1999.

\bibitem{meltzner2012persistent}
Aron~J Meltzner, Kerry Sieh, Hong-Wei Chiang, Chuan-Chou Shen, Bambang~W
  Suwargadi, Danny~H Natawidjaja, Belle Philibosian, and Richard~W Briggs.
\newblock {Persistent termini of 2004-and 2005-like ruptures of the {Sunda}
  megathrust}.
\newblock {\em Journal of Geophysical Research: Solid Earth}, 117(B4), 2012.

\bibitem{meltzner2010coral}
Aron~J Meltzner, Kerry Sieh, Hong-Wei Chiang, Chuan-Chou Shen, Bambang~W
  Suwargadi, Danny~H Natawidjaja, Belle~E Philibosian, Richard~W Briggs, and
  John Galetzka.
\newblock {Coral evidence for earthquake recurrence and an {AD} 1390--1455
  cluster at the south end of the 2004 {Aceh--Andaman} rupture}.
\newblock {\em Journal of Geophysical Research: Solid Earth}, 115(B10), 2010.

\bibitem{meltzner2015time}
Aron~J Meltzner, Kerry Sieh, Hong-Wei Chiang, Chung-Che Wu, Louisa~LH Tsang,
  Chuan-Chou Shen, Emma~M Hill, Bambang~W Suwargadi, Danny~H Natawidjaja, Belle
  Philibosian, et~al.
\newblock {Time-varying interseismic strain rates and similar seismic ruptures
  on the {Nias--Simeulue} patch of the {Sunda} megathrust}.
\newblock {\em Quaternary Science Reviews}, 122:258--281, 2015.

\bibitem{monecke20081}
Katrin Monecke, Willi Finger, David Klarer, Widjo Kongko, Brian~G McAdoo,
  Andrew~L Moore, and Sam~U Sudrajat.
\newblock {A 1,000-year sediment record of tsunami recurrence in northern
  {S}umatra}.
\newblock {\em Nature}, 455(7217):1232, 2008.

\bibitem{mulia2018adaptive}
Iyan~E Mulia, Aditya~Riadi Gusman, M~Jakir~Hossen, and Kenji Satake.
\newblock Adaptive tsunami source inversion using optimizations and the
  reciprocity principle.
\newblock {\em Journal of Geophysical Research: Solid Earth}, 123(12):10--749,
  2018.

\bibitem{nanayama2003unusually}
Futoshi Nanayama, Kenji Satake, Ryuta Furukawa, Koichi Shimokawa, Brian~F
  Atwater, Kiyoyuki Shigeno, and Shigeru Yamaki.
\newblock Unusually large earthquakes inferred from tsunami deposits along the
  kuril trench.
\newblock {\em Nature}, 424(6949):660--663, 2003.

\bibitem{newcomb1987seismic}
KR~Newcomb and WR~McCann.
\newblock {Seismic history and seismotectonics of the Sunda Arc}.
\newblock {\em Journal of Geophysical Research: Solid Earth}, 92(B1):421--439,
  1987.

\bibitem{nugroho2009plate}
Hendro Nugroho, Ron Harris, Amin~W Lestariya, and Bilal Maruf.
\newblock Plate boundary reorganization in the active banda arc--continent
  collision: Insights from new gps measurements.
\newblock {\em Tectonophysics}, 479(1-2):52--65, 2009.

\bibitem{okada1985surface}
Yoshimitsu Okada.
\newblock {Surface deformation due to shear and tensile faults in a
  half-space}.
\newblock {\em Bulletin of the seismological society of America},
  75(4):1135--1154, 1985.

\bibitem{okada1992internal}
Yoshimitsu Okada.
\newblock {Internal deformation due to shear and tensile faults in a
  half-space}.
\newblock {\em Bulletin of the Seismological Society of America},
  82(2):1018--1040, 1992.

\bibitem{okal2003mechanism}
Emile~A Okal and Dominique Reymond.
\newblock The mechanism of great banda sea earthquake of 1 february 1938:
  applying the method of preliminary determination of focal mechanism to a
  historical event.
\newblock {\em earth and planetary science letters}, 216(1-2):1--15, 2003.

\bibitem{OkalSynolakis2004}
Emile~A Okal and Costas~E Synolakis.
\newblock Source discriminants for near-field tsunamis.
\newblock {\em Geophysical Journal International}, 158(3):899--912, 2004.

\bibitem{owhadi2015brittleness}
Houman Owhadi, Clint Scovel, and Tim Sullivan.
\newblock On the brittleness of {B}ayesian inference.
\newblock {\em SIAM Review}, 57(4):566--582, 2015.

\bibitem{percival2011extraction}
Donald~B Percival, Donald~W Denbo, Marie~C Ebl{\'e}, Edison Gica, Harold~O
  Mofjeld, Michael~C Spillane, Liujuan Tang, and Vasily~V Titov.
\newblock Extraction of tsunami source coefficients via inversion of dart buoy
  data.
\newblock {\em Natural hazards}, 58(1):567--590, 2011.

\bibitem{PranantyoCummins2020}
Ignatius~Ryan Pranantyo and Phil~R Cummins.
\newblock The 1674 ambon tsunami: Extreme run-up caused by an
  earthquake-triggered landslide.
\newblock {\em Pure and Applied Geophysics}, 177(3):1639--1657, 2020.

\bibitem{reid2016two}
Anthony Reid.
\newblock {Two hitherto unknown Indonesian tsunamis of the seventeenth century:
  Probabilities and context}.
\newblock {\em Journal of Southeast Asian Studies}, 47(1):88--108, 2016.

\bibitem{roberts1997weak}
Gareth~O Roberts, Andrew Gelman, Walter~R Gilks, et~al.
\newblock {Weak convergence and optimal scaling of random walk Metropolis
  algorithms}.
\newblock {\em The annals of applied probability}, 7(1):110--120, 1997.

\bibitem{roberts1998optimal}
Gareth~O Roberts and Jeffrey~S Rosenthal.
\newblock {Optimal scaling of discrete approximations to Langevin diffusions}.
\newblock {\em Journal of the Royal Statistical Society: Series B (Statistical
  Methodology)}, 60(1):255--268, 1998.

\bibitem{rubin2017highly}
Charles~M Rubin, Benjamin~P Horton, Kerry Sieh, Jessica~E Pilarczyk, Patrick
  Daly, Nazli Ismail, and Andrew~C Parnell.
\newblock Highly variable recurrence of tsunamis in the 7,400 years before the
  2004 indian ocean tsunami.
\newblock {\em Nature communications}, 8:16019, 2017.

\bibitem{saito2011tsunami}
Tatsuhiko Saito, Yoshihiro Ito, Daisuke Inazu, and Ryota Hino.
\newblock Tsunami source of the 2011 tohoku-oki earthquake, japan: Inversion
  analysis based on dispersive tsunami simulations.
\newblock {\em Geophysical Research Letters}, 38(7), 2011.

\bibitem{sallares2019}
Valent{\'\i} Sallar{\`e}s and C{\'e}sar~R. Ranero.
\newblock Upper-plate rigidity determines depth-varying rupture behaviour of
  megathrust earthquakes.
\newblock {\em Nature}, 576(7785):96--101, 2019.

\bibitem{sieh2008earthquake}
Kerry Sieh, Danny~H Natawidjaja, Aron~J Meltzner, Chuan-Chou Shen, Hai Cheng,
  Kuei-Shu Li, Bambang~W Suwargadi, John Galetzka, Belle Philibosian, and
  R~Lawrence Edwards.
\newblock {Earthquake supercycles inferred from sea-level changes recorded in
  the corals of west {S}umatra}.
\newblock {\em Science}, 322(5908):1674--1678, 2008.

\bibitem{sraj2014uncertainty}
Ihab Sraj, Kyle~T Mandli, Omar~M Knio, Clint~N Dawson, and Ibrahim Hoteit.
\newblock Uncertainty quantification and inference of manning’s friction
  coefficients using dart buoy data during the t{\=o}hoku tsunami.
\newblock {\em Ocean Modelling}, 83:82--97, 2014.

\bibitem{sraj2017quantifying}
Ihab Sraj, Kyle~T Mandli, Omar~M Knio, Clint~N Dawson, and Ibrahim Hoteit.
\newblock Quantifying uncertainties in fault slip distribution during the
  t{\=o}hoku tsunami using polynomial chaos.
\newblock {\em Ocean Dynamics}, 67(12):1535--1551, 2017.

\bibitem{stuart2010inverse}
Andrew~M Stuart.
\newblock {Inverse problems: a {B}ayesian perspective}.
\newblock {\em Acta Numerica}, 19:451--559, 2010.

\bibitem{newspaper}
Jacob Swart.
\newblock {Verhandelingen en Berigten Betrekkelijk het Zeewegen en de
  Zeevaartkunde (English: Treatises and Reports Related to the Seaways and
  Nautical Sciences)}.
\newblock 13:257--274, 1853.

\bibitem{tanioka1996fault}
Yuichiro Tanioka and Kenji Sataka.
\newblock Fault parameters of the 1896 sanriku tsunami earthquake estimated
  from tsunami numerical modeling.
\newblock {\em Geophysical Research Letters}, 23(13):1549--1552, 1996.

\bibitem{tarantola2005inverse}
Albert Tarantola.
\newblock {\em {Inverse Problem Theory and Methods for Model Parameter
  Estimation}}.
\newblock siam, 2005.

\bibitem{tate2015australia}
Garrett~W Tate, Nadine McQuarrie, Douwe~JJ van Hinsbergen, Richard~R Bakker,
  Ron Harris, and Haishui Jiang.
\newblock {Australia going down under: Quantifying continental subduction
  during arc-continent accretion in Timor-Leste}.
\newblock {\em Geosphere}, 11(6):1860--1883, 2015.

\bibitem{turcotte2002geodynamics}
Donald~L Turcotte and Gerald Schubert.
\newblock {\em Geodynamics}.
\newblock Cambridge university press, 2002.

\bibitem{wells1994new}
Donald~L Wells and Kevin~J Coppersmith.
\newblock {New empirical relationships among magnitude, rupture length, rupture
  width, rupture area, and surface displacement}.
\newblock {\em Bulletin of the seismological Society of America},
  84(4):974--1002, 1994.

\bibitem{zenodov1}
Jared~P. Whitehead.
\newblock {tsunamibayes: 1852 paper submission}, October 2020.

\bibitem{whitford1977geochemistry}
DJ~Whitford, W~Compston, IA~Nicholls, and MJ~Abbott.
\newblock {Geochemistry of late Cenozoic lavas from eastern Indonesia: Role of
  subducted sediments in petrogenesis}.
\newblock {\em Geology}, 5(9):571--575, 1977.

\bibitem{wichmann1918earthquakes}
A~Wichmann.
\newblock {The earthquakes of the Indian archipelago to 1857}.
\newblock {\em Verhandl. Koninkl. Akad. van Wetenschappen, 2nd sec},
  20(4):1--193, 1918.

\bibitem{wichmann1922earthquakes}
A~Wichmann.
\newblock {The earthquakes of the Indian archipelago from 1858 to 1877}.
\newblock {\em Verhandl. Koninkl. Akad. van Wetenschappen, 2nd sec},
  22(5):1--209, 1922.

\end{thebibliography}

\end{document}